\documentclass[pdflatex,11pt]{article}
\usepackage{jheppub}
\usepackage{comment}
\usepackage{xcolor}
\usepackage{graphicx}
\usepackage{wrapfig}
\usepackage{multirow}

\newcommand{\ali}[1]{\begin{align} #1 \end{align}}
\newcommand{\p}{\partial}

\newcommand{\ra}{\rightarrow}

\newcommand{\vev}[1]{\langle #1 \rangle} 
\newcommand{\mn}{{\mu\nu}}

\begin{document}
	\thispagestyle{empty}

	\title{Radial canonical AdS$_3$ gravity and \boldmath $T\bar{T}$} 
	
	\author[a]{Matthew J. Blacker}\author[b]{, Nele Callebaut}\author[b]{, Blanca Hergueta}\author[c]{ and Sirui Ning}
	\affiliation[a]{Department of Applied Mathematics and Theoretical Physics,
		University of Cambridge, Cambridge CB3 0WA, UK}
	\emailAdd{mjb318@cam.ac.uk, nele.callebaut@thp.uni-koeln.de, hergueta@thp.uni-koeln.de, sirui.ning@physics.ox.ac.uk}
	\affiliation[b]{Institute for Theoretical Physics, University of Cologne, Z\"{u}lpicher Stra\ss e 77, 50937 K\"{o}ln, Germany}
	\affiliation[c]{The Rudolf Peierls Centre for Theoretical Physics, University of Oxford,\\Oxford OX1 3PU, UK}

	\abstract{ 
		We employ an ADM deparametrization strategy to discuss the radial canonical formalism of asymptotically AdS$_3$ gravity. It  
		leads to the identification of a radial `time' before quantization, which is the volume time, canonically conjugate to York time.  
		Holographically, 
		this allows to interpret the semi-classical 
		path integral of $T\bar T$ theory 
		as a Schr\"odinger wavefunctional satisfying a Schr\"odinger evolution equation in volume time, and the $T\bar T$ operator expectation value in terms of the Hamiltonian that generates 
		volume time translations -- both consistent with cut-off holography. 
		We make use of the canonical perspective to construct the rotating BTZ solution from the Hamilton-Jacobi equation, with a finite cut-off energy spectrum that has a known holographic $T\bar T$ interpretation, 
		as well as semi-classical Wheeler-DeWitt states for that solution.
	}
	
	\maketitle

	
	\section{Introduction and summary of results}

There has been recent interest in the canonical approach to quantum gravity in Anti-de Sitter (AdS) universes, e.g.~\cite{PapadoulakiChowdhury:2021nxw, Witten:2022xxp,WallAraujo-Regado:2022gvw}. The program of canonical quantization starts from the Hamiltonian formulation of GR, which allows one to identify configuration variables and their conjugate momenta. In the original context of $(3+1)$-dimensional gravity with zero cosmological constant, these variables are given by the 3-metric of Cauchy slices and corresponding conjugate momenta. Their Poisson brackets are promoted to commutators, and the classical Hamiltonian constraint promoted to a Wheeler-DeWitt (WdW) equation in what is known as `quantum geometrodynamics' or WdW quantum gravity. 
Applying  
this approach to quantum gravity to theories of gravity with negative cosmological constant is important to formulate the fundamental problems of AdS quantum gravity, which has been extensively - but somewhat indirectly - studied in the past two decades through holography. Especially in low dimensions, where AdS holography has  
proven to be very powerful, it 
seems possible to combine insights from WdW and holographic approaches \cite{Yang:2018gdb,Harlow:2018tqv,Maldacena:2019cbz}.

Another development, 
in the context of AdS holography, has been that of $T\bar T$ theory \cite{Smirnov:2016lqw,Cavaglia:2016oda}. This is the theory obtained from deforming 2-dimensional quantum field theories with Zamolodchikov's irrelevant $T\bar T$ operator \cite{Zamolodchikov:2004ce}. The resulting theory has non-local features but at the same time a closed form expression for its energy spectrum. This notion of integrability has led to many complementary perspectives on $T\bar T$ theory  \cite{Jiang:2019epa} other than that of non-local QFT, 
such as 2-dimensional gravity \cite{Dubovsky:2017cnj,Dubovsky:2018bmo,Callebaut:2019omt,Tolley:2019nmm}, random geometry \cite{Cardy:2018sdv}, and worldsheet string theory \cite{Cavaglia:2016oda,McGough:2016lol,Callebaut:2019omt} descriptions, which make $T\bar T$ theory a very rich theory to study. 
Moreover, when the undeformed theory is a 2-dimensional holographic CFT, the $T\bar T$ deformation also has a powerful 
holographic interpretation as `pulling the boundary inwards' 
or `cut-off holography' \cite{McGough:2016lol} (see also \cite{Guica:2019nzm,Marolf18,Caputa:2020lpa}),   
which has opened the door to a strategic approach to beyond AdS holography \cite{SilversteinGorbenko:2018oov,SilversteinBatra:2024kjl}. 

One of the 
main arguments for the holographic $T\bar T$ proposal is the connection 
between the $T\bar T$ flow and bulk evolution described by a radial WdW equation. 
It makes use of 
the long-known observation that the WdW equation in the semi-classical limit reduces to a Hamilton-Jacobi equation 
(which has been previously interpreted holographically in terms of holographic RG \cite{deBoer:1999tgo,Verlinde:1999xm,Heemskerk:2010hk,Freidel:2008sh}).  
In this way, $T\bar T$ forms a bridge between different quantum gravity approaches, namely WdW gravity on one hand and AdS/CFT on the other. In some sense, cut-off holography tells us that $T\bar T$ `dissects' AdS/CFT, 
and as such 
we can expect $T\bar T$ theory to provide a 
language to discuss and address issues in WdW AdS gravity. The real challenge will be to make statements beyond the semi-classical approximation.

The relation between $T\bar T$ and WdW gravity has been investigated in recent work \cite{Witten:2022xxp,WallAraujo-Regado:2022gvw}, which includes the proposal of Cauchy slice holography. In these papers, one discusses the standard canonical formalism in the context of AdS gravity, 
treating actual time as the direction of WdW evolution.

In this paper, in contrast, we study the \emph{radial} canonical formalism for AdS$_3$ (see also \cite{Freidel:2008sh}), which 
has been shown in \cite{McGough:2016lol} to have a direct $T\bar T$ interpretation at the level of the Hamilton-Jacobi equation and cut-off AdS energy spectrum. 
We revisit both these $T\bar T$ interpretations (Section \ref{sect2TTbar} and \ref{sectenergy}), in the context of our more detailed discussion of the canonical treatment. To this end, we construct the BTZ solution as well as corresponding WdW states by exponentiating solutions to the Hamilton-Jacobi equation, which form a semi-classical basis of the bulk theory. 
This follows the program of the recent works \cite{Hartnoll:2022snh,Blacker:2023oan,BlackerNing:2023ezy} (in $3+1$ dimensions), but applied to $2+1$ dimensions. In \cite{Hartnoll:2022snh,BlackerNing:2023ezy}, the constructed WdW states (upon including a counterterm contribution) were conjectured to be holographically equivalent to the Lorentzian partition function of a 
field theory that was left unspecified (but a possible higher-dimensional $T\bar T$ interpretation was mentioned in \cite{Hartnoll:2022snh}). We apply the same strategy to $2+1$ dimensions, precisely to explicitly identify the dual field theory as $T\bar T$ theory, which is 2-dimensional.

In Section \ref{sect2}, on the radial canonical formalism, we introduce a mini-superspace approximation that allows the identification of a preferred time before quantization: the volume time $v \equiv \sqrt{-\gamma}$, equal to the volume density of radial slices of the asymptotically AdS manifold. This time variable is canonically dual to the York time \cite{York1:1972sj}, 
which has been previously discussed in the context of AdS/CFT, e.g.~in \cite{Belin:2018bpg,Belin:2020oib,Witten:2022xxp}. 
The identification allows a reduced phase space formalism discussion of radial canonical AdS$_3$ gravity. 
In it, the Hamiltonian constraint takes the form of a Schr\"odinger equation for the WdW wavefunctional. 
`True' or physical canonical degrees of freedom are identified as the ones introduced in \cite{McGough:2016lol}, and the role of  unconstrained (i.e.~non-vanishing), 
`true' 
Hamiltonian density is taken by the York time $\pi_v$. At the semi-classical level, these statements have a $T\bar T$ interpretation. In particular, the semi-classical $T\bar T$ 
path integral solves a radial Schr\"odinger equation into the AdS bulk, and the 
value of the generator of volume time translations is given by the integrated expectation value of the $T\bar T$ operator $\vev{\mathcal O_{T\bar T}}$. 
We comment 
on the differences at the full quantum level, and on the role of the counterterm in the gravitational action in this discussion. 
The main result of  Section \ref{sect2} is the volume time Schr\"odinger equation \eqref{Schrodeq}, with semi-classical $T\bar T$ interpretation \eqref{ZQFT} and \eqref{HADMTTBar}. 
 
In Section \ref{sect3} we proceed to construct the BTZ solution \eqref{BTZsol} from the Hamilton-Jacobi equation. 
In particular, we include the technical complication of allowing rotating BTZ solutions, with an eye on the $T\bar T$ interpretation in which the rotation 
takes the role of momentum. 
Corresponding semi-classical solutions of the WdW equation are subsequently  constructed in Section \ref{sectsc}. The evolution of the obtained Vilenkin wavefunctional \eqref{EqOnShellWavefunction} is illustrated in the Penrose diagrams in Figures \ref{figBTZnonrot} and \ref{figBTZrot}.  
We discuss different measures of energy of the solution, and their holographically dual interpretation, in Section \ref{sectenergy}.

In Appendix \ref{AppADM} we provide some more background on the ADM deparametrization strategy \cite{Arnowitt:1962hi}, and in Appendix \ref{Appvol} we repeat the BTZ solution construction of Section \ref{sect3} in the volume notation employed in \cite{Hartnoll:2022snh}.

A comment on notation: throughout the paper we use a notation that distinguishes between objects in the theory without counterterm contribution to the action, and with counterterm. For reference, we summarize that notation here. In the presence of the counterterm, we use $\pi^{ij}$ for the canonical momenta, $S$ for the gravitational action, $S_{cl}$ for the on-shell action, and $\mathcal S$ for the Hamilton's principal function. In the absence of the counterterm, we use $p^{ij}$ for the canonical momenta, $I$ for the gravitational action, $I_{cl}$ for the on-shell action, and $\mathcal I$ for the Hamilton's principal function. They are simply related by \eqref{actionI} and \eqref{cantranf}.

In the final stages of preparing this paper, the work \cite{Godet:2024ich} appeared, 
which also employs a volume time interpretation of $T\bar T$ evolution, applied to de Sitter, as well as \cite{Kaushal:2024xob} on the notion of emergent time in  Hamiltonian General Relativity.

\section{Volume time reduced phase space} \label{sect2} 
	
We consider a $D = (2+1)$-dimensional theory of gravity with negative cosmological constant $\Lambda = -1/l^2$ 
in terms of AdS radius $l$. The gravitational action for a manifold $\mathcal M$ with $d=(D-1)$-dimensional timelike boundary $\p \mathcal M$ 
is given by   	
\ali{
	S &= \frac{1}{2\kappa} \int_{\mathcal M} d^D x \sqrt{-g} (R^{(D)} - 2 \Lambda) -\frac{1}{\kappa} \int_{\p \mathcal M} d^{D-1} x \sqrt{-\gamma} \, K    -\frac{1}{\kappa \, l} \int_{\p \mathcal M} d^{D-1} x \sqrt{-\gamma}  \label{Sgrav}
}	
with gravitational constant $\kappa = 8 \pi G$, 3-curvature $R^{(3)}$, trace of extrinsic curvature $K$, and the determinant of the 3-metric $g_\mn$ and induced 2-metric $\gamma_{ij}$ on $\p \mathcal M$ 
denoted $g$ and $\gamma$.  
The extrinsic curvature of $\p \mathcal M$ in $\mathcal M$ is defined as $K_\mn = -\frac{1}{2} \mathcal L_n \gamma_\mn$ for $\gamma_\mn = g_\mn - n_\mu n_\nu$ in terms of the normal $n^\mu$ to the boundary. We follow the same conventions as in \cite{Brown:1992br,Brown:1994gs}.

The first two terms in \eqref{Sgrav} are the Einstein-Hilbert action and Gibbons-Hawking-York boundary term, ensuring a well-defined variational principle. 
The last term is the standard counterterm $S_{ct} = \int d^{D-1} x \mathcal L_{ct} = -\frac{1}{\kappa \, l} \int d^{D-1} x \sqrt{-\gamma}$ \cite{Balasubramanian:1999re}. It is included to obtain a finite gravitational stress tensor. 

We write the spacetime metric in a $D$-dimensional ADM metric form 
\ali{
	ds^2 = N^2 dr^2 + \gamma_{ij}(dx^i + N^i dr) (dx^j + N^j dr)  \label{ADMmetric}
}
where we choose the decomposition in the radial direction, meaning we think of $r$ as playing the role of a (Euclidean) `time'. It describes a foliation with 
$d=2$-dimensional, timelike hypersurfaces $\Sigma$ homeomorphic to the boundary $\p \mathcal M$. We use Latin indices for the 2-coordinates $x^i$ and 2-metric $\gamma_{ij}$ on those constant $r$ slices.  
In the ADM decomposition, all functions are still functions of all coordinates, meaning lapse $N$, shift $N^i$ and 2-metric $\gamma_{ij}$ are short for $N(r,\vec x), N^i(r,\vec x), \gamma_{ij}(r,\vec x)$.

Using the Gauss-Codazzi equation for the ADM split of the curvature 
into 2-curvature $R$ and extrinsic curvature terms (e.g.~(A20) of \cite{Brown:1992br}), 
the action $S = \int dr \, d^2 x \, \mathcal L$   
becomes 
\ali{
	S &= \frac{1}{2\kappa} \int_{\mathcal M} d^D x \sqrt{-g} (R + K^2 - K^{ij} K_{ij} - 2 \Lambda) - \frac{1}{\kappa \, l} \int_{\p \mathcal M} d^{D-1} x \sqrt{-\gamma} . 
}
This expression can be brought in canonical form 
\ali{
	S &= \int_{\mathcal M} d^3 x \left( \pi^{ij}\p_r \gamma_{ij} - N \mathcal H - N^i \mathcal H_i \right)  , \label{action}
} 
in terms of the momenta 
\ali{
	\pi^{ij} \equiv \frac{\p \mathcal L}{\p \dot \gamma_{ij}} 
	= -\frac{1}{2\kappa} \sqrt{-\gamma}\, (K \gamma^{ij} - K^{ij} + \frac{1}{l} \gamma^{ij})  \label{defmom} \\
	p^{ij} \equiv \frac{\p \mathcal L}{\p \dot \gamma_{ij}} - \frac{\p \mathcal L_{ct}}{\p \dot \gamma_{ij}}= -\frac{1}{2\kappa} \sqrt{-\gamma}\, (K \gamma^{ij} - K^{ij})  \label{defmomp}
} 
where the dot refers to a derivative with respect to $r$, 
the Hamiltonian density 
\ali{
	\mathcal H &= -2 \kappa \,  G_{ijkl} \, p^{ij} p^{kl} - \frac{\sqrt{-\gamma}}{2\kappa} (R - 2 \Lambda) \nonumber \\ 
	&= -\frac{\sqrt{-\gamma}}{2\kappa} \, (R - 2 \Lambda) - \frac{2 \kappa}{\sqrt{-\gamma}} \left( p^{ij} p_{ij} - p^2 \right) , \label{defH}
} 
and momentum density 
\ali{
	\mathcal H_i = -2 D_j p_i^j. \label{defHi}
} 
The Hamiltonian density is given in terms of 
the DeWitt metric $G_{ijkl} = \frac{1}{2\sqrt{-\gamma}}  (\gamma_{ik} \gamma_{jl} + \gamma_{il} \gamma_{jk} - \frac{2}{d-1} \gamma_{ij} \gamma_{kl})$, and $D_j$ in \eqref{defHi} is the induced covariant derivative for tensors tangent to the boundary; 
traces of the momenta $p^k_k$ and $\pi^k_k$ are denoted $p$ and $\pi$. The DeWitt metric is the metric on superspace, which is the space of all 2-metrics $\gamma_{ij}$. 
The counterterm contribution to \eqref{action} enters through the last term of $\pi^{ij}$ in \eqref{defmom}. The densities $\mathcal H$ and $\mathcal H_i$ are given 
in their more standard form in terms of the momenta $p^{ij}$ without this counterterm contribution. 

The 2-metric and momenta in \eqref{action} form a pair of canonically conjugate variables $(\gamma_{ij}, \pi^{ij})$.    
The lapse and shift 
act as Lagrange multipliers imposing respectively the Hamiltonian and momentum constraints, $\mathcal H = 0$ and $\mathcal H_i = 0$. 
These constraints are the $G^r_\mu = 0$ Einstein field equations (rather than $G^t_\mu = 0$ in standard ADM slicing), 
so technically we could refer to $\mathcal H = 0$ as \emph{radial} Hamiltonian constraint \cite{Freidel:2008sh}, but we will not always repeat this specification explicitly, for reasons of conciseness. Written out in terms of $(\gamma_{ij},\pi^{ij})$, the (radial) Hamiltonian constraint is 
\ali{
	\frac{(2 \kappa)^2}{\gamma} \left( \pi_{ij} \pi^{ij} - \pi^2 \right) + \frac{4\kappa}{l} \frac{1}{\sqrt{-\gamma}} \,  \pi  - R(\gamma_{ij}) = 0 .   \label{pitotH}  
}
The effect of using a notation that includes the counterterm contribution 
is twofold: the appearance of a linear term in the momenta, and the disappearance of the cosmological constant term, compared to \eqref{defH}. 

Next, we make a symmetric ansatz for the metric in which all metric fields depend only on the radial coordinate 
\ali{
	N = N(r), \quad N^i = N^i(r), \quad \gamma_{ij}= \gamma_{ij}(r) . \label{minisup}
}
This has two immediate consequences. The first is that the momentum constraint $\mathcal H_i = 0$ is trivially satisfied. (Note that this would already be the case if only the hypersurface metric 
was restricted, $\gamma_{ij}= \gamma_{ij}(r)$, which is the radial equivalent of the `spatially constant' ansatz used in the canonical description of $(2+1)$-dimensional gravity in \cite{Martinec:1984fs}.) An ansatz with this property can be called a mini-superspace ansatz, with `mini' referring to the strong reduction in covered superspace, and this is indeed how we will refer to \eqref{minisup}.   
In the mini-superspace ansatz for the slicing of spacetime, the fact that the hypersurfaces are imposed to be non-intersecting means the Hamiltonian and momentum constraint 
are expressing diffeomorphism invariance of the spacetime. In general, such an interpretation of the constraints is more subtle \cite{Kiefer:2004xyv} (p.156). 

A second consequence of the mini-superspace ansatz 
is that the radial slicing is flat, 
more precisely $R \equiv R(\gamma_{ij})$ vanishes up to possible delta function contributions. With this form of the 2-curvature in mind, we will retain its presence in the equations. 
After the freezing of degrees of freedom \eqref{minisup}, 
the remaining freedom in the description is the choice of radial coordinate $r$ or `time'.

At the quantum level, the momenta are promoted to operators or 
variational derivatives \cite{Feng:2017xsh,Kiefer:2004xyv} 
\ali{
	\pi^{ij} 
	= -i \frac{\delta}{\delta \gamma_{ij}}  
	\label{Qmark}
}
when acting on the gravitational wavefunction(al) in a `coordinate representation' ($\psi$ depending on $\gamma_{ij}$)   
\ali{
	\psi \equiv \psi_\Sigma(\gamma) \,\,
	=  \int_{g|_{_{\Sigma}} = \gamma} 
	\mathcal D g \,\, e^{i S[g]}.  \label{psigrav}  
}
The Hamiltonian constraint $\mathcal H = 0$ is then promoted to the WdW equation 
\ali{ 
	\mathcal H \, \psi = 0, \label{WdW}
} 
with $\mathcal H$ now in the role of an operator. It expresses a restriction on physically allowed wavefunctionals $\psi$. 
Explicitly, it has the form 
\ali{
	\left( \frac{(2 \kappa)^2}{\gamma} \left( \pi_{ij} \pi^{ij} - \pi^2 \right) + \frac{4\kappa}{l} \frac{1}{\sqrt{-\gamma}} \,  \pi  - R(\gamma_{ij}) \right)  \psi = 0 . \label{WdWeq} 
}
The term that is quadratic in the momentum operators comes with a prescribed normal ordering \cite{halliwell_introductory_2009}.

In a semi-classical approximation, the wavefunction taking the WKB form $\psi \sim \exp{ i S_{cl} }$ and taking the limit $\kappa \ra 0$, 
one recovers from the WdW equation the Hamilton-Jacobi (HJ) equation  
\ali{
	\frac{(2 \kappa)^2}{\gamma} \left( -\frac{\delta S_{cl}}{\delta \gamma^{ij}} \frac{\delta S_{cl}}{\delta \gamma_{ij}} - \left(\gamma_{ij} \frac{\delta S_{cl}}{\delta \gamma_{ij}}\right)^2 \right) + \frac{4\kappa}{l} \frac{1}{\sqrt{-\gamma}} \, \gamma_{ij} \frac{\delta S_{cl}}{\delta \gamma_{ij}} - R(\gamma_{ij})   = 0 
}
with the on-shell action $S_{cl}$ in the role of Hamilton's principal function $\mathcal S$ \cite{Papadimitriou:2016yit}. 
It is the Hamiltonian constraint \eqref{pitotH}  
rewritten (using $\delta S_{cl} = \int \pi^{ij} \delta \gamma_{ij}$ on a solution) as a condition on the functional form of the on-shell action's dependence on the boundary metric, $S_{cl}[\gamma_{ij}]$.

Where the Hamiltonian density ${\mathcal H}$  
is purely quadratic in  
$p^{ij}$ in \eqref{defH}, it contains a term linear in the (trace of) the momentum $\pi^{ij}$, since $\pi^{ij} = p^{ij} - \frac{1}{2 \kappa l} \sqrt{-\gamma} \gamma^{ij}$. Therefore the linear term in the WdW equation can be traced back to the presence of the counterterm, i.e.~the fact that the action $S$ in the exponent of the wavefunction $\psi$ in \eqref{WdWeq} is the total gravitational action \emph{including} the counterterm.

The linear term in the WdW equation can now be identified as a variation with respect to the volume of radial slices. 
This follows from 
\ali{
	\pi = -i \, \gamma_{ij} \frac{\delta}{\delta \gamma_{ij}} = -i \sqrt{-\gamma} \frac{\delta}{\delta  \sqrt{-\gamma}} \,\, ,  
}   
which upon introducing the notation 
\ali{
	v \equiv \sqrt{-\gamma} \,\, , \qquad \pi_v \equiv -i \frac{\delta}{\delta  \sqrt{-\gamma}}  
}
for the volume of the radial slices and its canonical partner,  
can be written as 
\ali{
	\pi = v \, \pi_v .  \label{pipiv}
}   
This suggests a natural orthogonal decomposition of the momenta, obtained by splitting off the trace part $\pi$ or $\pi_v$ from a remaining traceless part,  
\ali{
	\pi^{ij} = \frac{\tilde \pi^{ij}}{\sqrt{-\gamma}} + \frac{1}{2} \gamma^{ij} \pi , \label{orthdec}
} 
and a rewriting of the 2-metric by splitting off the volume 
\ali{
	\gamma_{ij} = \sqrt{-\gamma} \,  \tilde \gamma_{ij}.  \label{gammatilde}
}
Or yet, 
\ali{
	\pi^{ij} = \frac{\tilde \pi^{ij}}{v} + \frac{1}{2} \tilde \gamma^{ij} \pi_v \,\, , \qquad \gamma_{ij} = v \,  \tilde \gamma_{ij}.  \label{newdof}
} 

The new introduced variables are $\tilde \pi^{ij}$, 
which has the property that it is traceless, and $\tilde \gamma_{ij}$, 
which has the property that its determinant equals minus one. 
The indices of these objects are raised and lowered by the new metric $\tilde \gamma_{ij}$ and its inverse (defined through $\tilde \gamma_{ik}\tilde \gamma^{kj} = \delta_i^j$). Therefore, $\pi_{ij} = \sqrt{-\gamma} \, \tilde \pi_{ij} + \frac{1}{2} \gamma_{ij} \pi$ and $\gamma^{ij} = \frac{1}{\sqrt{-\gamma}} \tilde \gamma^{ij}$. 
The tracelessness of $\tilde \pi^{ij}$ implies that the decomposition \eqref{orthdec} is indeed orthogonal as $\int d^2 x \, \frac{\tilde \pi^{ij}}{\sqrt{-\gamma}} \, \gamma_{ij} \pi = \int d^2 x \,  \tilde \pi \, \pi = 0$ \cite{York2:1973ia}.    
We've rediscovered the same variables that Martinec introduced in \cite{Martinec:1984fs}  for the description of canonical (2+1)-dimensional $\Lambda = 0$ gravity, and that more generally are 
employed in $D$-dimensional 
$\Lambda = 0$ canonical gravity discussions \cite{Kiefer:2004xyv,Hawking:1983hn} 
to argue that the volume of the slices forms the direction of time of the $(- + ... +)$ signature DeWitt metric $G_{ijkl}$ on $\frac{D(D-1)}{2}$-dimensional superspace. 
We will now see that in the case of radial slicing they lead to a reduced phase space description of (2+1)-dimensional $\Lambda < 0$ gravity.

In terms of the new variables, 
\eqref{pitotH} becomes 
\ali{
	-\frac{(2 \kappa)^2}{v^2} \tilde \pi^{ij} \tilde \pi^{kl} \tilde \gamma_{ki} \tilde \gamma_{lj} + \frac{(2 \kappa)^2}{2} \pi_v^2  + \frac{4\kappa}{l} \pi_v  - R(v,\tilde \gamma_{ij}) 
	= 0 \, . \label{Hamconstraintnew}
}
It is the Hamiltonian constraint $\mathcal H=0$ with the general $(\gamma_{ij}, \pi^{ij})$ dependence rewritten in terms of the orthogonal decomposition  $(v, \pi_v, \tilde \gamma_{ij}, \tilde \pi^{ij})$.  
Separating the linear term, it takes the form 
\ali{
	\pi_v   = \mathcal P^0(\pi_v, \, v, \, \tilde \pi^{ij}, \, \tilde \gamma_{ij} ). \label{ADM415}
} 
This can be solved for $\pi_v$ to 
\ali{
	\pi_v   = \mathcal T^0_0(v, \, \tilde \pi^{ij}, \, \tilde \gamma_{ij} )  \label{ADM416}
} 
with the explicit solution of the quadratic equation simply 
\ali{
	\mathcal T^0_0 \equiv -\frac{1}{\kappa l} + \frac{1}{\kappa l} \sqrt{1 + 2  \kappa^2 l^2 \,  \frac{\tilde \pi^{ij} \tilde \pi_{ij}}{v^2} + \frac{l^2}{2} R(v,\tilde \gamma_{ij})}  \, . \label{T00sol}
}
As we will shortly explain, the above reasoning is an ADM `deparametrization of gravity' argument. To make the comparison to the ADM notation of \cite{Arnowitt:1962hi} explicit, we have introduced in equations \eqref{ADM415}-\eqref{ADM416} the same notation as in that original paper.

What the 
rewriting to \eqref{ADM416} achieves, is that the Hamiltonian constraint -- rather than expressing the vanishing of the constrained Hamiltonian density $\mathcal H$ -- takes the more familiar form of a momentum equaling a `true' Hamiltonian density given by 
\ali{
	\mathcal H_{ADM} \equiv -\mathcal T^0_0 (v, \, \tilde \pi^{ij}, \, \tilde \gamma_{ij} ) .  \label{HADM}
}
The momentum necessary for this identification is the \emph{volume time} momentum $\pi_v$. 
Upon quantization, with $\pi_v = -i \, \delta/\delta v$, the Hamiltonian constraint then takes the form of the Schr\"odinger equation 
\ali{
	i \frac{\delta}{\delta v} \, \psi  =  H_{ADM} \, \psi. \label{Schrodeq}
	} 
The sign of the square root in the solution \eqref{T00sol} for $\mathcal T^0_0$ is chosen on the basis of boundedness of the Hamiltonian, and $H_{ADM} = \int d^2 x \, \mathcal H_{ADM}$ is the radial `true' Hamiltonian. 

The above argument follows the `deparametrization of gravity' strategy\footnote{As reviewed in \cite{Kiefer:2004xyv}, criticisms of this strategy can be kept in mind, e.g.~in \cite{Torre:1992rg} for the case of a closed universe.} 
of  Arnowitt, Deser and Misner \cite{Arnowitt:1962hi} and of Kuchar in e.g.~\cite{Kuchar:1971xm}.  
In mechanics, a system of $M$ degrees of freedom with action $I = \int dt \left( p_i \p_t q_i - H_{true}(p,q,t) \right)$ (with $i = 1, ..., M$) can be brought into a parametrized form $I = \int d\tau \,  p_j \p_\tau q_j$  $= \int d\tau \left( p_j \p_\tau q_j - N H \right)$ (with $j = 1, ..., M+1$) by writing Hamiltonian and time 
as a conjugate pair of variables 
$(t,-H_{true}) \equiv (q_{M+1}(\tau),p_{M+1}(\tau))$, with arbitrary parameter $\tau$ and new degree of freedom $q_{M+1}$. The second rewriting of $I$  
introduces a Lagrange multiplier $N(\tau)$ to impose the constraint equation $H(p_{M+1},q_{M+1},p_i, q_i) = 0$, which can be any equation with the solution $p_{M+1} = -H_{true}(p_i,q_i,q_{M+1})$.  
To reverse back from a parametrized to an unparametrized description\footnote{This is detailed in Appendix \ref{AppADM}.}, one should substitute the solution $p_{M+1} = -H_{true}(p_i,q_i,q_{M+1})$ of the constraint equation $H=0$, 
\emph{and} impose a `coordinate condition' $q_{M+1} \equiv t = \tau$ identifying the time $t$. In comparison to mechanics, the 
gravitational action \eqref{action} evaluated on 
the mini-superspace ansatz \eqref{minisup} comes in an `already parametrized' form $S = \int dr \, d^2 x \, (\pi^{ij} \p_r \gamma_{ij} - N \mathcal H)$, with the momentum constraint $\mathcal H_i \equiv 0$ already identically satisfied and `time' $r$ in the role of parameter. 
It can be similarly brought into a deparametrized form by substituting the solution of the constraint equations, with solution \eqref{ADM416} the equivalent of $p_{M+1} = -H(p_i,q_i)$, \emph{and} imposing the identification of time 
\ali{
	v \,  l = - r \,\, \qquad \text{(volume time)},  \label{voltime}  
}
where the AdS radius $l$ enters for reasons of dimensionality, and 
the sign is such that the infinite `past' is situated at the asymptotic AdS boundary. 
This volume time\footnote{
	Technically, $v$ is of course the volume \emph{density} of the $\Sigma$ slices, but we will refer to it as volume time.  
} identification $\sqrt{-\gamma} \, l = -r$ is thus 
the explicit equivalent of the coordinate condition $q_{M+1} = \tau$. The deparametrization results in singling out a choice of time before quantization. In the process, the number of degrees of freedom decreases (from $M+1$ to $M$ in mechanics) and the `true' degrees of freedom 
are observed to be the variables  
$\tilde \gamma_{ij}$ and their conjugate $\tilde \pi^{ij}$ on which the `true' or ADM Hamiltonian density in \eqref{HADM} depends. In the case at hand, $\mathcal H_{ADM}$ also depends on the volume time $v$ itself.

Let us now discuss the deparametrized form of the action, applying the ADM reduction technique of substituting the constraints and imposing coordinate conditions. 
Such a treatment of $(2+1)$-dimensional $\Lambda = 0$ gravity can be found in \cite{Moncrief:1989dx}, and indeed the following discussion for the $\Lambda < 0$ case will be very similar. 
Upon substituting \eqref{orthdec} and \eqref{gammatilde}, the action \eqref{action} evaluated on the constraints becomes 
\ali{
	S 
	&= \int d^3 x \left[  \left(  \frac{\tilde \pi^{ij}}{v} + \frac{1}{2} \gamma^{ij} \pi \right) \left( v \, \p_r \tilde \gamma_{ij} + \tilde \gamma_{ij}\p_r v\right)  \right] \\ 
	&= \int dr \, d^2 x \,\,( \tilde \pi^{ij} \, \p_r \tilde \gamma_{ij} + \pi_v \, \p_r v   ). 
	\label{actionnewvar}
}
Going to the second line, the cross terms disappear due to the properties of the new variables $\tilde \gamma_{ij}$ and $\tilde \pi^{ij}$, that is,  $\sqrt{-\tilde \gamma} = 1$ 
and $\tilde \pi = 0$. After the identification of volume time 
\eqref{voltime}, we find 
\ali{
	S = \int dv \, d^2 x \,\,( \tilde \pi^{ij} \, \p_v \tilde \gamma_{ij} + \pi_v ) 
	}  
for the reduced phase space action. It indeed takes the form from the mechanics example, $I = \int dt \left( p_i \p_t q_i - H_{true}(p,q,t) \right)$, such that we can read off from it that 
the true Hamiltonian density is given (on the constraint) by the conjugate partner of volume time   
\ali{
	\mathcal H_{ADM} = - \pi_v .  
	\label{HADMeq} 
} 
This is of course consistent with \eqref{ADM416} and \eqref{HADM}. 
The corresponding true Hamiltonian is obtained by integrating over the boundary directions $H_{ADM} = -\int d^2 x \, \pi_v/l$. 

The momentum $\pi_v$ is known in the literature 
as \emph{York time} \cite{York1:1972sj}. 
From \eqref{defmom} and \eqref{pipiv}, the momenta have trace $\pi = -\frac{1}{2\kappa} \sqrt{-\gamma} (K + \frac{2}{l})$ and 
\ali{ 
	\pi_v =  -\frac{1}{2\kappa} \left( K + \frac{2}{l} \right).  
}     
The second term is the contribution from the counterterm, which is not present in the flat gravity context in which York time was introduced \cite{York1:1972sj}. In that case, York time is just proportional to the trace of the extrinsic curvature (of spacelike slices in that context), but it is still the conjugate of volume, with the `true' Hamiltonian in \cite{York1:1972sj, Moncrief:1989dx} identified as the  
volume of the universe. 
ADM deparametrization in terms of York time has been discussed in \cite{Moncrief:1989dx} and in the context of cosmology in \cite{Roser:2014foa}.

\subsection{$T\bar T$ flow equation interpretation}  \label{sect2TTbar}

The radial HJ equations 
have been holographically interpreted as $T\bar T$ trace flow equations in \cite{McGough:2016lol} and \cite{Marolf18} (in Euclidean signature). 
We will repeat their argument here (in Lorentzian signature), showing that 
the volume time identification provides a natural, useful notation.

Given a classical solution, variations of the on-shell action $S_{cl}$ (or `on-shell variations' \cite{Banados:2016zim}) take the form \cite{Brown:1992br} 
\ali{
	\delta S_{cl} &= \int_\Sigma  d^2 x \, \pi^{ij} \delta \gamma_{ij}.  \label{deltaScl0}
} 
We rewrite this in terms of the new canonical pairs in our radial canonical gravity discussion, given by the volume time and its canonical partner $(v, \pi_v)$ and the `true' canonical degrees of freedom $(\tilde \gamma_{ij}, \tilde \pi^{ij})$, to the simple form 
\ali{
	\delta S_{cl} = \int_\Sigma  d^2 x \, \left(  \tilde \pi^{ij} \delta \tilde \gamma_{ij} + \pi_v \, \delta v \right) .  \label{deltaScl} 
} 
The absence of cross terms again follows from the same argument as in the derivation of \eqref{actionnewvar}. 
From \eqref{deltaScl0}, we have $\pi^{ij} = \frac{\delta S_{cl}}{\delta \gamma_{ij}}$ and from the second form in \eqref{deltaScl}, 
\ali{
	\tilde \pi^{ij} = \frac{\delta S_{cl}}{\delta \tilde \gamma_{ij}} \qquad \text{and} \qquad \pi_v = \frac{\delta S_{cl}}{\delta v} . \label{momofScl}
	}
The solution's gravitational energy-momentum 
that is contained in the bulk region bounded by the constant $r$ surface $\Sigma$ with intrinsic metric $\gamma_{ij}$ is  
given by the Brown-York stress tensor \cite{Brown:1992br} 
\ali{
	T^{ij} = \frac{2}{\sqrt{-\gamma}} \frac{\delta S_{cl}}{\delta \gamma_{ij}} \label{BYdef}
}
or $T_{ij} = -\frac{2}{\sqrt{-\gamma}} \frac{\delta S_{cl}}{\delta \gamma^{ij}}$. It is immediate that the stress tensor is proportional to the momenta $T^{ij} = \frac{2}{\sqrt{-\gamma}} \pi^{ij}$. 
From a $T\bar T$ perspective, we are specifically interested in the 
combination of Brown-York stress tensor components (with notation $\Theta$ for the trace $T_k^k$) 
\ali{
	T_{ij} T^{ij} - \Theta^2   
} 
which written out in terms of the canonical degrees of freedom is 	
\ali{ 
	\frac{4}{v^2} \left( \pi_{ij} \pi^{ij} - \pi^2 \right) 
	= 4 \left( \frac{\tilde \pi_{ij} \tilde \pi^{ij}}{v^2} - \frac{1}{2} \pi_v^2 \right) . \label{OTTbarofpiv}
}
This is indeed the combination of momenta that also appears in the Hamiltonian constraints \eqref{pitotH} and \eqref{Hamconstraintnew}, the latter more precisely taking the form  
\ali{
	\pi_v = \kappa l \,  \left( \frac{\tilde \pi_{ij} \tilde \pi^{ij}}{v^2} - \frac{1}{2} \pi_v^2 \right)   \label{HJtocompare} 
}
(with vanishing curvature $R$). 
With \eqref{momofScl}, these 
Hamiltonian constraints    
give HJ equations for $S_{cl}[\gamma_{ij}]$ or $S_{cl}[\tilde \gamma_{ij},v]$.

Now let us consider $T\bar T$ theory, a 2-dimensional QFT defined on a background metric $\gamma_{ij}$ obtained by deforming a seed theory, which in our considerations will always be a holographic 2-dimensional CFT. 
Following the notation 
of \cite{Tolley:2019nmm}, 
this `$T\bar T$ theory' or `$T\bar T$-deformed CFT' in Lorentzian signature 
is described by an action 
that satisfies 
\ali{
	\frac{\p}{\p \lambda} S^{(\lambda)}_{QFT} = \frac{1}{2} \int d^2 x \, \sqrt{-\gamma} \, \mathcal O_{T\bar T}  \label{LorTTbar} 
} 
with $S^{(0)}_{QFT}$ the seed theory action, and $S^{(\lambda+d\lambda)}_{QFT}$ the perturbatively deformed one from $S^{(\lambda)}_{QFT}$. 
The deformation is given by the $T\bar T$ operator
\ali{
	\mathcal O_{T\bar T} \equiv T^{QFT}_{ij} T_{QFT}^{ij} - (\Theta_{QFT})^2
}
which is well-defined as a composite local operator 
up to derivatives of other local operators \cite{Zamolodchikov:2004ce},   
and where $T^{QFT}_{ij}$ is the stress tensor of the deformed theory $S^{(\lambda)}_{QFT}$, and $\Theta_{QFT}$ its trace. 
The $T\bar T$ coupling $\lambda$ is the parameter that is positive in the 2-dimensional massive gravity description of $T\bar T$ \cite{Tolley:2019nmm} and in the Nambu-Goto string description of $T\bar T$ \cite{Cavaglia:2016oda,McGough:2016lol,Dubovsky:2017cnj,Callebaut:2019omt},  
being respectively identified with the inverse mass squared of the graviton and the Nambu-Goto string tension. It has dimension length squared, hence the deformation is irrelevant. 

Following the argument of \cite{Marolf18,Jiang:2019epa} 
that in the absence of other scales in the theory (apart from the role of the UV cutoff which is less clear), $\mu \, dS^{(\lambda)}_{QFT}/d\mu = \int d^2 x \sqrt{-\gamma}\,  \Theta_{QFT}$ for mass scale $\mu = 1/\sqrt{\lambda}$, the so-called trace flow equation of $T\bar T$ follows:   
\ali{
	\vev{\Theta_{QFT}} = -\lambda \vev{\mathcal O_{T\bar T}}. \label{traceflow}
}

In regular AdS/CFT, the Brown-York stress tensor $T^{ij}$ is holographically dual to the stress tensor expectation value $\vev{T^{ij}_{CFT}}$ of the CFT living on the conformal boundary. The cut-off holographic duality 
of \cite{McGough:2016lol} 
proposes that similarly, the Brown-York stress tensor $T^{ij}$ of the bulk solution cut off at a \emph{finite} value of the radial direction 
is holographically dual to the stress tensor expectation value $\vev{T^{ij}_{QFT}}$ of a $T\bar T$ theory living on the slice $\Sigma$ with metric $\gamma_{ij}$.  
Applying this prescription to the trace flow equation, the left hand side in bulk notation becomes 
the trace of the Brown-York stress tensor $\Theta = \frac{2}{v} \pi = 2 \pi_v$, and the right hand side $(-\lambda)$ times \eqref{OTTbarofpiv}, 
such that the bulk translation of \eqref{traceflow} reads 
\ali{
	\pi_v &= - 2 \lambda \left( \frac{\tilde \pi_{ij} \tilde \pi^{ij}}{v^2} - \frac{1}{2} \pi_v^2 \right).  
}
Comparison to the radial HJ equation \eqref{HJtocompare} leads to exact agreement, provided one makes the identification\footnote{Our $\lambda$ relates to $\mu$ in \cite{McGough:2016lol} as $\mu = 4 \lambda$.} 
\ali{
	\lambda &= -\frac{1}{2} \kappa l    \label{lambdaVMM}
} 
between the QFT parameter $\lambda$ and gravitational parameters $\kappa \equiv 8 \pi G$ and $l$.  
The sign is such that the $T\bar T$ coupling has the opposite sign in a holographic context, sometimes referred to as holographic $T\bar T$ having the `wrong sign', compared to in e.g.~the Nambu-Goto interpretation. 
In a `hybrid' (meaning left hand side in bulk and right hand side in boundary) notation, we could write 
\ali{
	\pi_v = -\frac{\lambda}{2} \vev{\mathcal O_{T\bar T}} . 
} 
Comparing the trace flow equation in this holographic notation to \eqref{HADMeq}, we can identify the expectation value of the $T\bar T$ operator  with the on-shell value of the ADM Hamiltonian (density) generating evolution in the 
$v$-direction into the bulk, 
\ali{
	\frac{1}{2} \lambda \vev{\mathcal O_{T\bar T}} = l \, \mathcal H_{ADM} . \label{HADMTTBar} 
}
This observation in terms of volume time fits naturally in the cut-off holographic idea that $T\bar T$ deformations push the boundary inwards. 

To arrive at the $\lambda$ identification \eqref{lambdaVMM}, first obtained in \cite{McGough:2016lol}, we have essentially reproduced their argument. This includes the decomposition of metric components and momenta in \eqref{orthdec} and \eqref{gammatilde}, which was already employed in that paper. Our discussion 
adds a canonical gravity viewpoint on a natural origin for this decomposition. Namely, as discussed in Section \ref{sect2}, it is the decomposition that allows an ADM deparametrization of the classical GR problem for a special choice of `time'. The volume time 
identification then leads to a simple notation, starting with \eqref{deltaScl}, for the derivation of \eqref{lambdaVMM}, and a natural interpretation of the $T\bar T$ operator expectation value in terms of the 
corresponding ADM Hamiltonian.  

The same $\lambda$ identification \eqref{lambdaVMM} follows directly from comparing the trace flow equation to the constraint \eqref{pitotH} in terms of the original momenta, which is the approach of \cite{Marolf18}.  
Let us remark that the relation between the $T\bar T$ coupling $\lambda$ and volume time $v$   
can be made more explicit as follows. Similar to e.g.~in \cite{Hartman:2018tkw,Guica:2019nzm}, we can alternatively interpret the dual $T\bar T$ theory as living on the background geometry $\tilde \gamma_{ij}$, as defined in \eqref{gammatilde}, rather than $\gamma_{ij}$. 
This gives rise to the inversely proportional relation   
\ali{
	\lambda &= -\frac{1}{2} \frac{\kappa l}{v} .   \label{lambdaHK}
} 
We will however in the remainder of this paper stick to the interpretation of the $T\bar T$ theory on the induced boundary metric $\gamma_{ij}$, and correspondingly use \eqref{lambdaVMM}.

Modulo Weyl anomaly subtleties, the $T\bar T$ flow equation interpretation of the radial HJ equation implies that the radial WdW equation for the semi-classical gravitational wavefunctional $\psi$ can be interpreted as a $T\bar T$ flow equation for the semi-classical, Lorentzian $T\bar T$ partition function $Z_{QFT} \approx e^{i S_{QFT}}$ \cite{McGough:2016lol}.  
Said otherwise, at the semi-classical level, $\psi$ is holographically dual to $Z_{QFT}$, or in hybrid notation  
\ali{
	\psi = Z_{QFT}.  \label{ZQFT}
}
Since in Section \ref{sect2} we have reinterpreted the radial WdW equation as a volume time Schr\"odinger equation \eqref{Schrodeq}, we immediately 
arrive at the conclusion 
that the semi-classical, Lorentzian $T\bar T$ partition function 
solves a radial Schr\"odinger equation describing volume time evolution into the holographic bulk. This statement is 
closely related to the holographic RG discussion in terms of radial Schr\"odinger equations in \cite{Heemskerk:2010hk}. 

We note that the presence of the counterterm $S_{ct}$ in the bulk dual played a role in arriving at a Schr\"odinger interpretation of the WdW equation. Indeed we emphasized in 
Section \ref{sect2} that it is the linear term in the momenta in the WdW equation that 
suggested a rewriting in Schr\"odinger form \eqref{Schrodeq}, and the origin of the linear term $\sim \pi_v$ is the counterterm in the bulk action. (It is worth emphasizing this, because the counterterm is no longer required to be present at a finite radial cut-off from the perspective of renormalization.)  
From an AdS/CFT perspective, the volume time appears because the asymptotic radial WdW equation for $\psi$ expresses the Weyl anomaly equation for the CFT path integral $Z_{CFT}$. 
The Weyl anomaly equation or conformal Ward identity describes the behavior of $Z_{CFT}$ under scale transformations, or, variations of the volume $v$. The corresponding classical statement is that the asymptotic Hamilton-Jacobi equation $\pi_v = \frac{l}{4 \kappa} R$  expresses the Weyl anomaly $T_\mu^\mu = \frac{c}{24 \pi} R$. We add that it does so most succinctly with volume time because $T_\mu^\mu$ equals $2 \pi_v$. The $T\bar T$ deformation then provides the holographic dual interpretation to the non-asymptotic HJ and WdW equations, i.e.~including quadratic terms in \eqref{Hamconstraintnew}, or from the boundary point of view, extending evolution along the RG flow direction.

The volume time interpretation 
provides a revealed structure of true canonical degrees of freedom and a preferred radial time on the bulk side, as well as 
an indication to the relevance of York time in the description of the system. Since volume time is dual to York time, the semi-classical $T\bar T$ 
path integral $\psi(\gamma_{ij})$ 
is related by a Fourier transform to the functional \cite{Belin:2020oib, Blacker:2023oan} 
\ali{ 
	\tilde \psi(\pi_v,\tilde \gamma_{ij}) &= \int dv \, \psi(v,\tilde \gamma_{ij}) \, e^{-i \int d^2 x \, v \, \pi_v} . 
}  
	The WdW equation for $\tilde \psi$ will describe evolution in York time rather than volume time (with minus $v$ providing the corresponding Hamiltonian density, similar to in \cite{York1:1972sj}). 
This perspective should allow to further investigate the role of the mixed boundary conditions discussed in \cite{Witten:2022xxp} (see also \cite{Anninos:2023epi,Anninos:2024wpy}). 
They consist of keeping $\pi_v$ and the conformal class of metrics on $\Sigma$ fixed, and are argued in \cite{Witten:2022xxp} to be 
better defined from a WdW 
equation point of view than the Dirichlet boundary conditions appearing in cut-off holography \cite{McGough:2016lol} (this is more pressing in higher-dimensional set-ups). We leave this for future work.

The appearance of volume time in mini-superspace AdS gravity is not unexpected. Firstly, the associated true canonical degrees of freedom in \eqref{newdof}  
are directly equivalent to the Dirac canonical variables discussed in \cite{Regge:1974zd} in $(3+1)$-dimensional $\Lambda = 0$ gravity. 
Secondly, the volume time is the natural Anti-de Sitter equivalent of the use of volume time in de Sitter quantum cosmology \cite{Kiefer:2004xyv,Hawking:1983hn,Godet:2024ich}, as the radial direction into the AdS bulk takes over the role of the time direction in dS \cite{Strominger:2001pn}. While in dS the volume will keep expanding, providing a monotonous notion of time, in AdS, the volume time will reach a maximum at the maximal volume slice. This is illustrated further on in Figure \ref{figvolBTZ}. From a $T\bar T$ perspective, this does not give rise to  complications, as the volume time provides a good dual description of $T\bar T$ in particular in the near-boundary limit in the bulk, or $c \ra \infty$ limit in the boundary theory. We discuss the reason for this regime in the next paragraph.

Going to the full quantum level, it is quite unclear if the Schr\"odinger wavefunction interpretation of the Lorentzian $T\bar T$ partition function provides any added value. 
This is because equations \eqref{Schrodeq} and \eqref{WdWeq} describe \emph{different} quantum gravity theories, obtained from different quantization (i.e.~normal ordering) prescriptions. The latter is described by the WdW equation with standard Laplace normal ordering \cite{halliwell_introductory_2009}, which is indeed the prescription followed in \cite{Hartnoll:2022snh,Blacker:2023oan,BlackerNing:2023ezy} and the one we will follow in Section \ref{sectsc}. 
The former is instead described by a WdW equation with normal ordering of the square root in \eqref{T00sol}. 
If one thinks of 
quadratic normal ordering in the trace flow equation $\Theta_{QFT} = -\lambda : \mathcal O_{T\bar T} :$ as a \emph{definition} of the quantum $T\bar T$ theory, then the normal ordering prescriptions of \cite{halliwell_introductory_2009} or \cite{WallAraujo-Regado:2022gvw, Witten:2022xxp} 
provide more natural bulk interpretations when interested in beyond semi-classical statements, as compared to the normal ordering prescription of the square root in \eqref{T00sol} in our volume time interpretation \eqref{Schrodeq}.

\begin{figure}
	\centering
	\includegraphics[width=6cm]{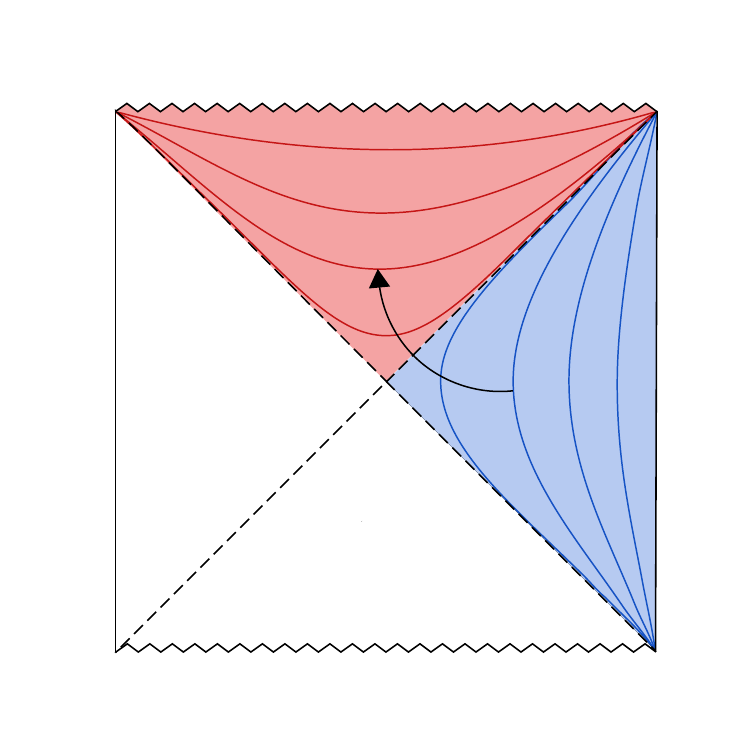} \qquad \qquad \qquad 
	\includegraphics[width=6cm]{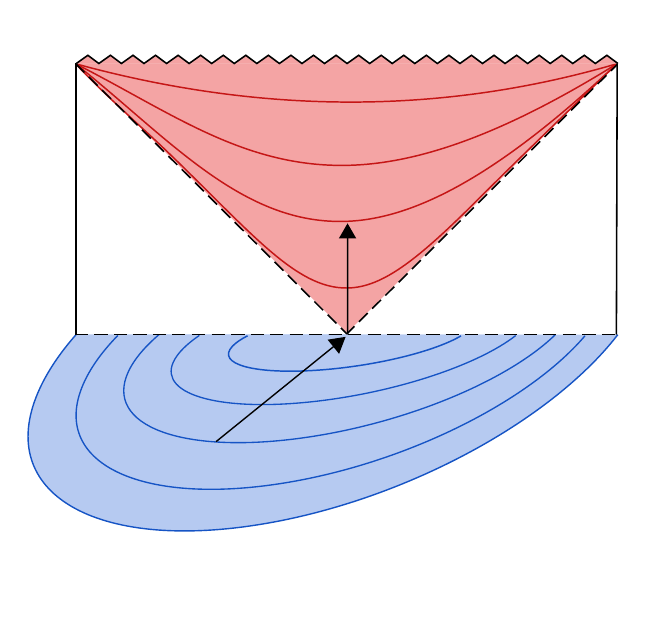} 
	\caption{Non-rotating BTZ solution. Left: radial evolution in Lorentzian signature, with constant-$R$ slices $\Sigma$ in the outer (blue) and inner (red) horizon region. Right: Euclidean radial evolution (blue) up to the horizon, followed by Lorentzian radial evolution (red).  
	} 
	\label{figBTZnonrot}
\end{figure}

\section{BTZ from the Hamilton-Jacobi equation} \label{sect3}

\begin{wrapfigure}{L}{0.29\textwidth}
	\centering
	\includegraphics[width=0.22\textwidth]{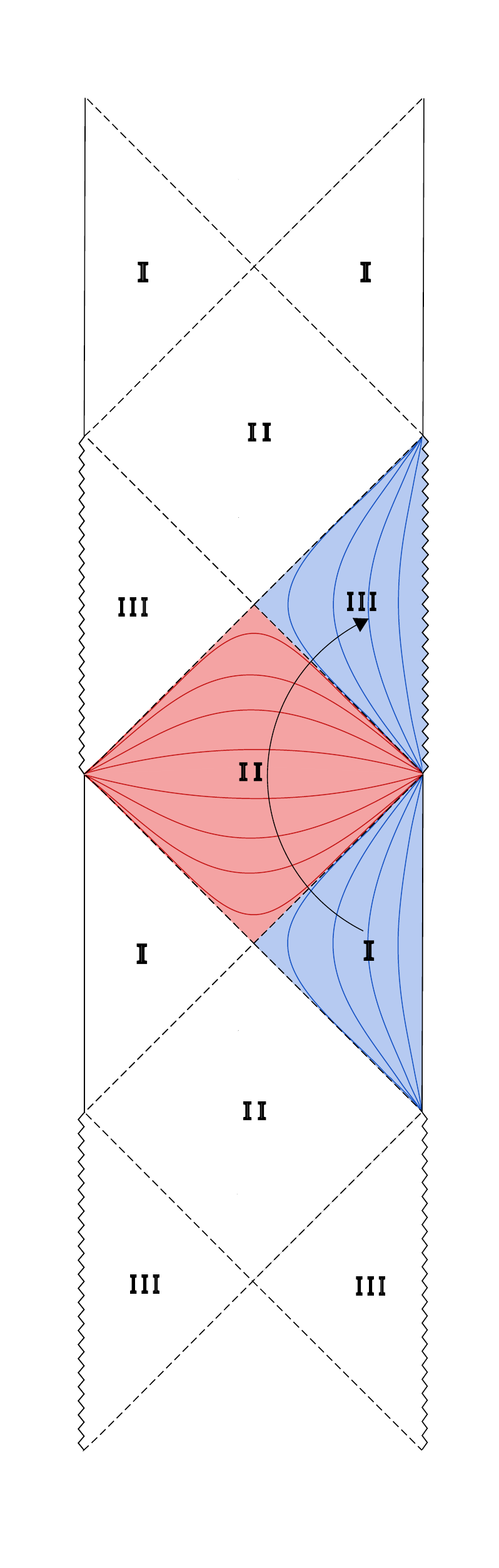} 
	\caption{Rotating BTZ solution. Radial evolution in the exterior (blue) and interior (red) regions, with constant-$R$ slices $\Sigma$ as solid lines. \\ \\ } 
	\label{figBTZrot}
\end{wrapfigure}

We have so far reinterpreted 
$T\bar T$ flow in terms of a radial WdW solution evolving in volume time. To make the connection between the $T\bar T$-deformed quantum theory and the WdW solution more explicit, it will be useful to consider a particular example and construct a map between quantities on each side of the duality. In particular, in this section we construct the BTZ solution from a canonical perspective. The corresponding Hamilton's principal function  
will be used in the next section to construct WdW states, and finally we will discuss the holographic interpretations of different energy measures for this solution. To construct the BTZ solution, we closely follow in this section the planar AdS$_4$-Schwarzschild solution derivation of \cite{Hartnoll:2022snh}, applied to one lower dimension. Working in 2+1 dimensions has two benefits. Firstly, it makes it possible to compare directly to holographic $T\bar T$ arguments \cite{McGough:2016lol}, 
without having to resort to proposed higher-dimensional extensions of $T\bar T$ theory \cite{Hartman:2018tkw,Taylor:2018xcy}. Secondly, it allows us to include the technical complication of rotation of the black hole. To our knowledge, the derivation of Kerr from the Hamilton-Jacobi function is e.g.~yet to be explored in the literature.  \\
	
To obtain the BTZ solution \cite{Banados:1992wn} from the HJ equation, we will start from the gravitational action \eqref{Sgrav} \emph{without} the counterterm  
\begin{equation}
	I = S - S_{ct},  \label{actionI}
\end{equation} 
i.e.~work with the Hamiltonian 
\eqref{defH} in terms of the momenta $p^{ij}$,  
and 
account for the counterterm's effect afterwards by the canonical (i.e.~Poisson bracket preserving) transformation $(\gamma_{ij},p^{ij})$ to $(\gamma_{ij}, \pi^{ij})$ with    
\ali{
	\pi^{ij} = p^{ij} - \frac{1}{2\kappa l} \sqrt{-\gamma} \gamma^{ij} . \label{cantranf}
} 
We consider 
the mini-superspace ansatz \eqref{minisup} introduced in the previous section, given by the ADM metric \eqref{ADMmetric} with $r$-dependent lapse and induced metric functions $N(r)$ and $\gamma_{ij}(r)$, now written as  
\begin{align} 
	ds^2 & = - N^2(r) dr^2 + \gamma_{ij}(r) dx^i dx^j \nonumber \\
	& = - N^2 dr^2 + \gamma_{tt} dt^2 + 2 \gamma_{t \varphi} dt d\varphi +  
	R^2 d\varphi^2  \label{EqAnsatz} 
\end{align}
with vanishing $N^i(r)$ (fixing the  
freedom of choosing coordinates on the slice)   
and $\gamma_{\varphi \varphi}(r) \equiv R^2(r)$.  
For the two coordinates on the $\Sigma$ slices we have chosen the notation $x^i = (t,\varphi)$ with angular coordinate $\varphi$ for a manifold  
$\mathbb R \times \mathbb R \times S^1$. 
Compared to \eqref{ADMmetric}, the sign of $dr^2$ was changed to consider solutions in the region where $r$ is actually timelike (and $N^2$ positive), 
i.e.~the ansatz \eqref{EqAnsatz} will describe the black hole interior in Figures \ref{figBTZnonrot} and \ref{figBTZrot},  
where $r$ is the timelike coordinate. 

We can instead introduce a notation that allows to construct the solution in both the interior and exterior regions, by writing the ansatz 
\ali{ 
	ds^2 & = \epsilon(- N^2 dr^2 + \gamma_{tt} dt^2) + 2 \gamma_{t \varphi} dt d\varphi + 
	R^2 d\varphi^2 , \label{EqAnsatzbis} 
}
with $\epsilon = 1$  in the interior ($r$ timelike and $t$ spacelike) and $\epsilon = -1$ in the exterior ($r$ spacelike and $t$ timelike), for $N^2$ and $\gamma_{tt}$ assumed positive.  
A similar discussion can be found in \cite{BenAchour:2023dgj}.

On this ansatz \eqref{EqAnsatzbis}, we evaluate the action 
\ali{
	I &= \frac{1}{2\kappa} \int_{\mathcal M} d^3 x \sqrt{-g} \left(R^{(3)} + \frac{2}{l^2}\right) -\frac{1}{\kappa} \int_{\p \mathcal M} d^{2} x \sqrt{\gamma} \, K 
}
to be 
\begin{align}
	I\left[ N,\gamma_{ij} \right] &  = \int d^3 x \,\mathcal L , 
	\label{EqAction}
\end{align}
where the Lagrangian density 
is 
\begin{align}
	\mathcal L & = \frac{N}{\kappa l^2} \sqrt{|\epsilon \gamma_{tt}\gamma_{\varphi\varphi}-{\gamma_{t\varphi}}^2|} +  \frac{\left( \partial_r \gamma_{t\varphi} \right)^2 - \epsilon \left( \partial_r \gamma_{tt} \right) \left( \partial_r \gamma_{\varphi \varphi} \right)}{4\kappa N \sqrt{|\epsilon \gamma_{tt}\gamma_{\varphi\varphi}-{\gamma_{t\varphi}}^2|}}. 
	\label{EqLagrangian}
\end{align}
From this Lagrangian, we obtain the momenta $p^{ij}$ conjugate to the metric functions $\gamma_{ij}$ in the usual way \eqref{defmomp},   
and construct the Hamiltonian 
\begin{align}
	N \mathcal{H} = N \frac{\sqrt{|\epsilon\gamma_{tt} \gamma_{\varphi \varphi} - {\gamma_{t\varphi}}^2|}}{\kappa} \left[ \kappa^2 \left( (p^{t\varphi})^2 -  \epsilon \,  4 p^{tt} p^{\varphi \varphi} \right) - \frac{1}{l^2} \right]. 
	\label{EqHclassical}
\end{align}
The lapse $N$ is a Lagrange multiplier that imposes the Hamiltonian constraint $\mathcal{H} = 0$, or explicitly, 
\begin{align}
	&\frac{1}{\kappa^2 l^2} + \epsilon \, 4 \, p^{tt} p^{\varphi \varphi} - (p^{t\varphi})^2  = 0. 
	\label{EqHconstraint}
\end{align}
Upon writing the momenta as $p^{ij} = \partial_{\gamma_{ij}} \mathcal I(\gamma_{ij})$, per definition of the Hamilton's principal function $\mathcal I(\gamma_{ij})$, the Hamiltonian constraint becomes the Hamilton-Jacobi equation 
\begin{align}
	&\frac{1}{\kappa^2 l^2} - \left( \partial_{\gamma_{t\varphi}} \mathcal I \right)^2 + \epsilon \,  4 \left( \partial_{\gamma_{tt}} \mathcal I \right) \left( \partial_{\gamma_{\varphi \varphi}} \mathcal I \right) = 0 . 
	\label{EqHJequation}
\end{align}
There is a family of solutions to \eqref{EqHJequation}. We will in particular consider
\begin{align}
	&\mathcal I \left(\gamma_{ij};c,j \right) = - \epsilon \,  \frac{1}{4 \kappa^2 c} \gamma_{tt} + \frac{c}{l^2} \gamma_{\varphi \varphi} + \frac{\left( c j - \gamma_{t\varphi}/\kappa \right)^2}{4c \gamma_{\varphi \varphi}} . 
	\label{EqS}
\end{align}
Here $\lbrace c, j \rbrace$ are non-trivial constants of integration.  
Specifically, they are the Hamilton-Jacobi constants, with which we construct the phase space of our classical solution. 
A third constant of integration, contributing an overall shift in $\mathcal I$, could be included. Such a constant would not contribute to dynamics and is thus not considered further. There is also a freedom to choose the overal sign of $\mathcal I$.

In Hamilton-Jacobi theory, the solution to the classical equations of motion (the Euler-Lagrange equations for \eqref{EqLagrangian}) is obtained by introducing another pair of constants $\lbrace m, w \rbrace$ such that
\begin{align}
	&\partial_c \, \mathcal I = m \quad \text{ and } \quad \partial_j  \mathcal I = w . \label{HJcsts} 
\end{align}
Simultaneously solving these equations we can rewrite two of our metric functions in terms of the third. In particular, we obtain
\begin{align}
	&\gamma_{tt} = \epsilon \,  4 \kappa^2 \left( \gamma_{\varphi \varphi} \left( w^2 - \frac{c^2}{l^2} \right) + c \left( c \, m - j w \right) \right), \text{ and } \gamma_{t\varphi} = \kappa \left( c j - 2 w \gamma_{\varphi \varphi} \right). 
	\label{Eqgttgtphionshell}
\end{align}
We are now in a position to recover the geometry associated with this classical solution. 
We substitute \eqref{Eqgttgtphionshell} into the equation of motion for $N$ from \eqref{EqLagrangian}, to find 
\begin{align}
	N^2 dr^2 = \epsilon \, \frac{(d\gamma_{\varphi \varphi})^2}{4 m \gamma_{\varphi \varphi} - j^2 - 4\gamma_{\varphi \varphi}^2/l^2} . 
\end{align}
We now have all functions in \eqref{EqAnsatz} in terms of $\gamma_{\varphi \varphi} \equiv R^2$,  
so we can write for the (finally $\epsilon$-independent) metric solution  
\begin{align}
	\begin{split}
		ds^2 = & - \left( f_{m,j}(R) - \frac{\left( j - 2w R^2/c \right)^2}{4R^2} \right) d\left( 2 c\kappa t \right)^2 + \frac{1}{f_{m,j}(R)} dR^2 \\
		& + \left( j - 2 w R^2/c \right) d\varphi \, d \left( 2c\kappa t \right) + R^2 d\varphi^2 ,
	\end{split} 
	\label{EqMetricSolution}
\end{align}
where
\begin{align}
	f_{m,j}(R) = -m + \frac{j^2}{4R^2} + \frac{R^2}{l^2}.  \label{fmj}
\end{align}
We have thus recovered the BTZ black hole solution \cite{Banados:1992wn}. Explicitly, if we define the rotated angular coordinate and a rescaled time 
\begin{align} 
	& \phi  = -  \varphi + 2 w \kappa t, \text{ and } T = 2 c \kappa t 
	\label{EqPhiTtransform}
\end{align}
we can write \eqref{EqMetricSolution} in the more familiar form 
\begin{align}
	ds^2 = - f_{m,j}(R) dT^2 + \frac{1}{f_{m,j}(R)} dR^2 + R^2 \left( d{\phi} - \frac{j}{2R^2} d{T} \right)^2. \label{BTZsol} 
\end{align}
The transformation \eqref{EqPhiTtransform} preserves the periodicity of $\varphi$.

We have so far only considered one member of the family of solutions to \eqref{EqHJequation}. Provided that we do not use both elements of a conjugate pair $\lbrace c, m \rbrace$ or $\lbrace j, w \rbrace$, we could construct a solution using any combination of these constants. In particular, we could have considered the solution
\begin{align}
	\bar{\mathcal I}\left( \gamma_{tt},\gamma_{t\varphi},\gamma_{\varphi \varphi}; m, w \right) = \frac{1}{\kappa l} \sqrt{\gamma_{\varphi \varphi} - m l^2} \sqrt{- \epsilon \gamma_{tt} - 4 w \kappa \gamma_{t\varphi} - 4 w^2 \kappa^2 \gamma_{\varphi \varphi}}.
	\label{EqSbar}
\end{align}
The classical equations of motion are solved when
\begin{align}
	\partial_m \bar{\mathcal I} = c, \text{ and } \partial_w \bar{\mathcal I} = j, 
\end{align}
and we once again obtain \eqref{Eqgttgtphionshell}. 
We introduce this solution because it will emerge in the Fourier transform of Wheeler-DeWitt states in Section \ref{SecSemiclassicalStates}.  
It is important to emphasize that when evaluated on shell, \eqref{EqS} and \eqref{EqSbar} are equal to each other and the on-shell action (up to a constant which acts as a phase).

\begin{figure}[t]
	\centering 
	\includegraphics[width=7cm]{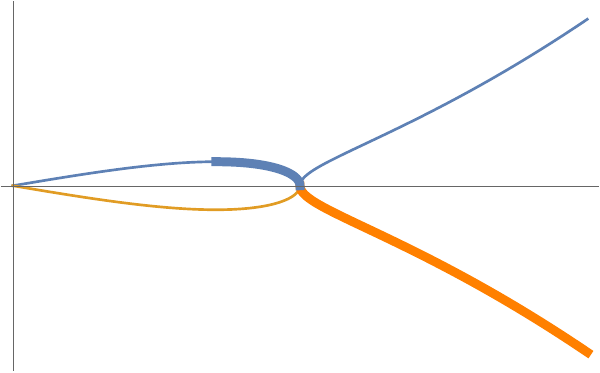} \qquad \includegraphics[width=7cm]{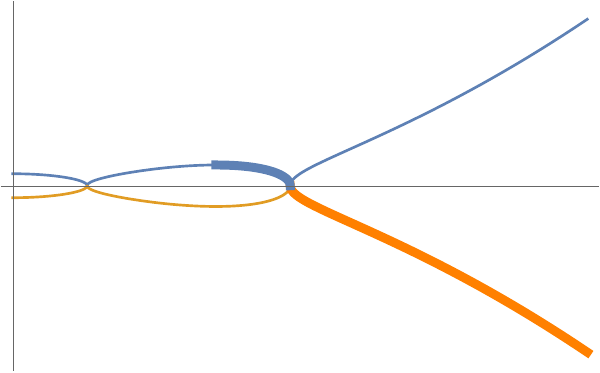}
	\caption{Illustrative plot of the volume density $v$ of radial BTZ slices (in blue), and $-v$ (in orange) as a function of radial coordinate $R$, 
		for choices of constants $(m, j, l) = (1,0,1)$ (left) and $(m, j, l) = (1,1/2,1)$ (right). 
		The volume time $r = \epsilon \, v \, l$ is highlighted in thick. It runs from the boundary to the outer horizon (orange thick) and then continues up until the interior maximal volume slice (blue thick).}   
	\label{figvolBTZ} 
\end{figure}

\paragraph{BTZ volume time} The volume time \eqref{voltime} for the BTZ  solution is 
\ali{
	r  = \epsilon R \sqrt{|f_{m,j}(R)|} = \epsilon R \sqrt{-\epsilon f_{m,j}(R)} \label{volumetimeBTZ}   
}
with $\epsilon = -1$ outside the outer horizon $R>r_+$ and $\epsilon = 1$ inside the outer horizon $r_- < R < r_+$. That is, radial `time' $r$ is equal to $r= -v l$ in the outside-region and to $r = v l$ in the inside-region (where we've rescaled by a factor $2 c \kappa l$ to write \eqref{volumetimeBTZ}). Defined in this way, and as illustrated in Figure \ref{figvolBTZ}, the volume time $r = \epsilon \, v \, l$ is monotonous, running from the boundary to the horizon, and then continuing until the interior maximal volume slice, which is located at a value of $R$ in the domain $r_- < R < r_+$. 
Beyond that, 
it no longer provides a well-defined, monotonous radial time.  
In the former region, which is the one we consider, constant  
$r \equiv \epsilon \, v \, l$ surfaces correspond to constant $R$ surfaces.  
The highlighted volume time evolution in Figure \ref{figvolBTZ} corresponds to the arrow of radial bulk evolution 
in Figure \ref{figBTZnonrot} for the non-rotating case, and in Figure \ref{figBTZrot} for the rotating case.

\section{Semi-classical Wheeler-DeWitt states} \label{sectsc}

We study the semi-classical WdW states corresponding to the classical BTZ solution discussed in the previous section. Note that in this section, we restrict the discussion to the interior, setting $\epsilon = 1$. It is straightforward to include $\epsilon$ back in.

It is worth noting that here we have parameterized our WdW states in a different way to in  \cite{Hartnoll:2022snh,Blacker:2023oan,BlackerNing:2023ezy}. There, the metric function $g_{tt}$ was used, and solutions  in one region of spacetime were continued into a causally disconnected region by extending the domain of $g_{tt}$ to be both positive \textit{and} negative. For instance, in \cite{Hartnoll:2022snh}, this was to allow a Lorentzian partition function at future infinity ($g_{tt} \rightarrow - \infty$) to prepare a wavefunction at the singularity ($g_{tt} \rightarrow + \infty$). In this work, we interpret the dual theory (the $T\bar T$-deformed CFT) to live on a slice in the same causal region as the WdW state. Thus, we use our $\epsilon$ to indicate which causal region that is. This has a more natural connection with our volume time interpretation of the WdW equation presented in Section \ref{sect2}.

\subsection{From classical to semi-classical}
\label{SecClassicaltoSemiclassic}
In order to construct semi-classical quantum states of the bulk, we follow the canonical approach of \cite{dewitt_quantum_1967}. 
The classical Hamiltonian is promoted to an operator, and the classical Hamiltonian constraint $\mathcal{H} = 0$ becomes a constraint on quantum states $\Psi$, namely the Wheeler-DeWitt equation  $\mathcal{H} \Psi = 0$ in  \eqref{WdW}.
Observe that in \eqref{EqAction} we have an action of the form 
\begin{align}
	&I[N, \gamma_{ij}] 
	= \int d^3 x \, N \left[ \frac{1}{2N^2}G_{AB} \dot{\gamma}^{A} \dot{\gamma}^{B} - V(\gamma_A) \right], 
	\label{Eqactiongeneral}
\end{align}
and in \eqref{EqHclassical} we have a corresponding total Hamiltonian of the form
\begin{align}
	&{
	N \mathcal{H} = N \left[ \frac{1}{2} G^{AB} p_{A} p_{B} + V(\gamma_A) \right]. }  \label{eq4dot2}
\end{align}
Here, $V(\gamma_A)$ is an effective potential of the metric functions $\gamma_{ij}$ and $G_{AB}$ is the mini-superspace DeWitt metric as in  \cite{dewitt_quantum_1967}. Each index $A,B$ corresponds to two spacetime indices $tt,t\varphi, \varphi \varphi$, 
with $\gamma_{ij}$ in the role of configuration space coordinates $q^A$ or $\gamma^A$, and $p^{ij}$ 
in the role of their corresponding momenta $p_A$. The $(A,B)$ notation, compared to the $(ijkl)$ notation in Section \ref{sect2}, follows the quantum cosmology literature conventions, see e.g.~chapter 8 of \cite{Kiefer:2004xyv}.

In constructing the operator ${\mathcal{H}}$, momenta are promoted to operators, and there is an ambiguity in the order of derivatives. The choice we consider here is the same as in \cite{Hartnoll:2022snh,Blacker:2023oan,BlackerNing:2023ezy}; which although differing from the recent work of \cite{WallAraujo-Regado:2022gvw,Witten:2022xxp}, follows the original prescription of \cite{hawking_operator_1986,halliwell_derivation_1988} and does not affect the leading order semi-classical physics. Specifically, we require  
that the quantization procedure should be covariant with respect to coordinate transformations in mini-superspace, or equivalently redefinitions of the metric functions. This can be achieved by choosing as the Wheeler-DeWitt equation \cite{hawking_operator_1986,halliwell_derivation_1988}
\begin{align}
	&{\left( - \frac{\hbar^2}{2} \nabla^2 + \hbar^2 \, \eta \, \mathcal R + V(\gamma_{ij}) \right) \Psi = 0. } 
	\label{EqWDWgeneral}
\end{align}
Here, $\nabla^2$ is the Laplacian for the  
DeWitt metric  
$\sqrt{-G}\nabla^2 = \partial_{\gamma^A} \left( \sqrt{-G} G^{AB} \partial_{\gamma^B} \right)$, $\mathcal R$ is the curvature of the DeWitt metric, and $\eta$ an arbitrary constant. The role of the $\eta$ term is to ensure invariance of the Wheeler De-Witt equation under lapse rescalings in the classical action \eqref{Eqactiongeneral}. Specifically, the Wheeler De-Witt equation is invariant under lapse rescalings $N \rightarrow \tilde{N} = \Omega^{-2} N$ if (see e.g.~\cite{halliwell_introductory_2009,Kiefer:2019bxk})   
\begin{align}
	\eta = - \frac{(n-2)}{8(n-1)},   
\end{align}
where $n$ is the dimension of the mini-superspace, for $n \geq 2$. Note that $\eta = 0$ for mini-superspaces with only two degrees of freedom, such as in \cite{Hartnoll:2022snh,Blacker:2023oan,BlackerNing:2023ezy}, so the $\eta$ term in \eqref{EqWDWgeneral} is then not present.  
In our present discussion, we are dealing with a three-dimensional mini-superspace; meaning that we require $\eta = -1/16$ for our WdW equation to be invariant under lapse rescalings \cite{halliwell_introductory_2009,halliwell_derivation_1988}. 

However, the presence of the $\eta$ term, and even the operator ordering ambiguity, are not relevant if one is only concerned with leading order semi-classical physics, as we are in this work. Observe that for solutions of the form $e^{i \mathcal I/\hbar}$, the leading order contribution ($\mathcal{O}(\hbar^0)$) is
\begin{align}
	&{\frac{1}{2} \left( \nabla \mathcal I \right)^2 + V(\gamma_{ij}) = 0, } 
\end{align}
which is the Hamilton-Jacobi equation for \eqref{Eqactiongeneral}. Indeed, the same leading order contribution is obtained for any choice of operator ordering. That is, in the semi-classical regime, it is possible to form a basis of Wheeler-DeWitt states by exponentiating the Hamilton-Jacobi solution.  

\subsection{Semi-classical states of the BTZ black hole}
\label{SecSemiclassicalStates}
For our set-up, we can read off that the mini-superspace DeWitt metric (equal to $G^{ijkl}$) is
\begin{align}
	G_{AB} = \frac{1}{4\kappa \sqrt{g_{tt}g_{\varphi \varphi} - {g_{t\varphi}}^2}} \begin{pmatrix}
		0 & 0 & -1 \\
		0 & 2 & 0 \\
		-1 & 0 & 0
	\end{pmatrix}
\end{align}
where $A,B$ refer to $tt, t\varphi, \varphi\varphi$. 
The Wheeler-DeWitt equation, with $\eta = -1/16$, is therefore
\begin{align}
	\begin{split}
		& {\left[ \hbar^2 \left( \frac{1}{4} \frac{\partial^2}{\partial \gamma_{t\varphi}^2} - \frac{\partial}{\partial \gamma_{tt}} \frac{\partial}{\partial \gamma_{\varphi \varphi}} \right) + \frac{1}{4 			\kappa^2 l^2} \right. }\\
		&{ \left. + \frac{ \hbar^2}{8\left( \gamma_{tt}\gamma_{\varphi \varphi} - \gamma_{t\varphi}^2 \right)} \left( \gamma_{\varphi \varphi} \frac{\partial}{\partial \gamma_{\varphi \varphi}} + 				\gamma_{tt} \frac{\partial}{\partial \gamma_{tt}} + \gamma_{t\varphi} \frac{\partial}{\partial \gamma_{t\varphi}} + \frac{3}{8} \right) \right] \Psi = 0. } 
	\end{split}
	\label{EqWDWequation}
\end{align}

From Section \ref{SecClassicaltoSemiclassic}, we know that to construct a basis of semi-classical solutions to \eqref{EqWDWequation}, we can consider ${e^{\pm \mathcal{I}/\hbar}}$, where ${\mathcal{I}}$ 
is a solution to \eqref{EqHJequation}. We therefore introduce the basis
\begin{align}
	\Psi(\gamma_{tt},\gamma_{t\varphi},\gamma_{\varphi \varphi};c,j) = e^{i\mathcal{I}(\gamma_{tt},\gamma_{t\varphi},\gamma_{\varphi \varphi};c,j)/\hbar}, 
	\label{EqPsiBasis}
\end{align}
where here ${\mathcal{I}}$  
is the solution introduced in \eqref{EqS}. The basis of solutions \eqref{EqPsiBasis} solves \eqref{EqWDWequation} up to leading semi-classical order, that is $\mathcal{O}(\hbar^0)$. From the basis \eqref{EqPsiBasis}, we construct a general solution 
\begin{align}
	&{\Psi \left( \gamma_{tt}, \gamma_{t\varphi},\gamma_{\varphi \varphi} \right) = \int_{-\infty}^{\infty} \frac{dc}{2\pi} \int_{-\infty}^{\infty} \frac{dj}{2\pi} \beta(c,j) e^{i\mathcal{I}(\gamma_{tt},\gamma_{t\varphi},\gamma_{\varphi \varphi};c,j)},} 
	\label{EqPsiGeneralcJ}
\end{align}
where ${\beta(c,j)}$ 
is an arbitrary function. Here, we have only considered the ${e^{+i \mathcal{I}}}$ 
solution in our sum. This is the Vilenkin choice of retaining only `outgoing modes' \cite{Vilenkin:1987kf, halliwell_introductory_2009}.  
It corresponds to the choice of red region that the $\Sigma$ slices transition into under WdW evolution, in Figures \ref{figBTZnonrot} and \ref{figBTZrot}, as the outer horizon is crossed.  
It is important to pick either the ${e^{+i \mathcal{I}}}$ or the ${e^{-i \mathcal{I}}}$ branch if one wishes to define a positive definite norm when introducing clocks (see \cite{Hartnoll:2022snh,BlackerNing:2023ezy} for further discussion). A superposition of the ${e^{\pm i \mathcal{I}}}$  solutions would lead to decoherence in the wavefunction, a phenomenon we are not concerned with in this work.  

We can obtain another set of semi-classical solutions by constructing a basis of states from \eqref{EqSbar}. That is, we could consider the basis
\begin{align}
	& {\Psi \left( \gamma_{tt},\gamma_{t\varphi},\gamma_{\varphi \varphi}; m, w \right) = e^{i\bar{\mathcal{I}} \left( \gamma_{tt},\gamma_{t\varphi},\gamma_{\varphi \varphi}; m, w \right)}.} 
	\label{EqPsiBasis2}
\end{align}
We can actually recover states formed from the basis \eqref{EqPsiBasis2} by Fourier transforming our solutions \eqref{EqPsiGeneralcJ}. We define $\alpha(m, w)$ by 
\begin{align}
	&{\beta(c,j)  = \int \int \frac{dm}{\sqrt{2\pi}} \frac{dw}{\sqrt{2\pi}} \alpha(m, w) e^{-icm -i w j}.}
	\label{Eqbasischange}
\end{align}
Substituting \eqref{Eqbasischange} into \eqref{EqPsiGeneralcJ} and evaluating the $c$ and ${j}$ integral by a 2D stationary phase approximation, we obtain 
\begin{align}
	&{\Psi \left( \gamma_{tt}, \gamma_{t\varphi},\gamma_{\varphi \varphi} \right) = \int_{-\infty}^{\infty} \frac{dm}{2\pi} \int_{-\infty}^{\infty} \frac{dw}{2\pi} \alpha(m,w) \frac{l\sqrt{\gamma_{\varphi \varphi}}}{\sqrt{\gamma_{\varphi \varphi} - ml^2}} \left( e^{i \left( \bar{\mathcal{I}} + \pi/4 \right)} + e^{-i\left( \bar{\mathcal{I}} + \pi/4 \right)}  \right).}
\end{align}
To semi-classical order, the prefactors are subleading; thus, the upshot of the above calculation is that we can alternatively construct general semi-classical solutions as
\begin{align}
	&{\Psi \left( \gamma_{tt}, \gamma_{t\varphi},\gamma_{\varphi \varphi} \right) = \int_{-\infty}^{\infty} \frac{dm}{2\pi} \int_{-\infty}^{\infty} \frac{dw}{2\pi} \alpha(m,w) e^{i \bar{\mathcal{I}}}.}
\end{align}
Again here, we have only considered the ${e^{+ i \bar{\mathcal{I}}}}$  branch of solutions. 

Of course, \eqref{Eqbasischange} is nothing more than a change of basis; we are yet to consider a particular form for ${\beta(c,j)}$. It is natural to consider Gaussian wavepackets, as these are strongly supported on the classical solution. Let us consider a wavepacket
\begin{align}
	\begin{split}
		{\beta(c,j) =} & {N_{\beta} \exp \left \lbrace - i \, m_0 \left( c - c_0 \right) - \frac{\Delta_c^2}{2} \left( c- c_0 \right)^2 \right \rbrace} \\
		& {\times \exp \left \lbrace - i \,w_0 j - \frac{\Delta_j^2}{2} \left( j- j_0 \right)^2 \right \rbrace.}\\
	\end{split} 
\label{EqBetaGaussian}
\end{align} 
Here, $N_{\beta} = \sqrt{\Delta_c \Delta_j/4\pi}$ 
is the normalisation of the wavepacket, and $\Delta_c,\Delta_j$ are its inverse width in the $c$ and $j$ directions respectively. These wavepackets are strongly peaked around ${\lbrace c = c_0, j = j_0 \rbrace}$ if ${\Delta_c, \Delta_j \gg 1}$. We can now explicitly compute \eqref{EqPsiGeneralcJ} for the wavepackets \eqref{EqBetaGaussian} by once again implementing a 2D stationary phase approximation. The wavefunction is strongly peaked on values of the metric function such that
\begin{align}
	&{\left. \frac{\partial \mathcal{I}}{\partial c} \right|_{c=c_0} = m_0, \text{ and } \left. \frac{\partial \mathcal{I}}{\partial j} \right|_{j = j_0} = w_0.}
\end{align}
These are of course the conditions which recover the classical solution \eqref{Eqgttgtphionshell} with ${m=m_0}$ and $w = w_0$. As discussed before, we only consider one of the stationary points in the stationary phase approximation. To leading semi-classical order, we therefore obtain
\begin{align}
	\begin{split}
		{\Psi \left( \gamma_{tt}, \gamma_{t\varphi}, \gamma_{\varphi \varphi}; c_0, j_0, m_0, w_0 \right) =} & {\delta \left( \gamma_{tt} = 4\kappa^2 \left( \gamma_{\varphi \varphi} \left( w^2 - \frac{c^2}{l^2} \right) + c \left( cm - jw \right) \right) \right)} \\
		& {\times \delta \left( \gamma_{t\varphi} = \kappa \left( cj - 2 \gamma_{\varphi \varphi} w \right) \right)} \\
		& {\times \exp \left \lbrace i c_0 \left( \frac{2\gamma_{\varphi \varphi}}{l^2} - m_0 \right) \right \rbrace} . 
	\end{split}
	\label{EqOnShellWavefunction}
\end{align}
The delta functions in \eqref{EqOnShellWavefunction} impose the classical solution \eqref{Eqgttgtphionshell}. Evaluating the on-shell action (that is, \eqref{EqAction} on the classical solution), we obtain ${2c_0 \gamma_{\varphi \varphi}/l^2}$. That is, the phase in \eqref{EqOnShellWavefunction} is simply the on-shell action, up to a phase which is a consequence of our choice of gaussian wavepacket \eqref{EqBetaGaussian}. 

\subsection{Quantum spread of the WdW states}  \label{sect43}
The semi-classical quantum wavefunction presented in \eqref{EqOnShellWavefunction} is highly localized upon the classical solution, and neglects any quantum spread in the wavefunction. In usual quantum mechanics, one may quantify the effect of quantum spread in the wavefunction by computing expectation values of physical observables. To do so, one requires a definition of an inner product, which is a conserved quantity in the Hilbert space of the quantum theory --- a quantity conserved with respect to some notion of time. That same notion of time describes the evolution of the quantum state $\Psi$ in some evolution equation (for example, the Schr\"{o}dinger equation). As we have emphasised, the Wheeler DeWitt equation encodes a relation between metric functions on a given slice of spacetime. In that sense, it is ``timeless''. Therefore, in order to define an inner product and compute expectation values, one needs to choose a coordinate (or some combination of coordinates) to treat as fixed, and label as a ``clock''. Changing the value of that clock is equivalent to evolving between slices. 

For an $(n+1)$-dimensional minisuperspace, if one chooses the $0$-th coordinate as a clock, one can use the DeWitt norm introduced in \cite{dewitt_quantum_1967} to define an inner product
\begin{align}
	|\Psi|^2_{\, q^0} \propto - \frac{i}{2} \int dq^1 \dots dq^n \left[ \Psi^* \left( \sqrt{-G} \,G^{0A} \partial_{A} 
	\Psi \right) - \text{h.c.} \right], \label{DeWittnorm}
\end{align}
where we employ the notation introduced under Eq.~\eqref{eq4dot2}. 
One has the freedom to rescale the norm appropriately. One natural choice of clock for our Wheeler DeWitt equation is $\gamma_{\varphi \varphi}$, as it is monotonic from the singularity to the boundary. 
In particular, this will make it straightforward to compute the quantum effects as one passes through the horizon. For the Wheeler DeWitt equation in \eqref{EqWDWequation}, the norm for the clock $\gamma_{\varphi \varphi}$ is 
\begin{align}
	|\Psi|^2_{\gamma_{\varphi\varphi}}  
	\propto & -\frac{i}{2} \int d\gamma_{tt} d\gamma_{t\varphi} \frac{1}{(\gamma_{tt}\gamma_{\varphi \varphi} - \gamma_{t\varphi}^2 )^{1/4}} \left( \Psi^* \partial_{\gamma_{tt}} \Psi - \text{h.c.} \right).
	\label{EqNormUnscaled} 
\end{align}
Because of the $(\gamma_{tt}\gamma_{\varphi \varphi} - \gamma_{t\varphi}^2 )^{-1/4}$ factor in the norm, completing computations with \eqref{EqNormUnscaled} is unwieldy for the basis of solutions we have proposed in \eqref{EqPsiGeneralcJ}, as they are not linear in this quantity. However, as discussed previously, Wheeler DeWitt equations related by a rescaling of the lapse function are physically equivalent. As such a rescaling redefines the DeWitt metric, it also redefines the inner product. Specifically, it introduces a factor of $\Omega^{2-n}$ into the integrand. We find that it is in particular convenient to introduce the lapse rescaling 
\begin{align}
	\Omega = \gamma_{\varphi \varphi} \left( \gamma_{tt} \gamma_{\varphi \varphi} - \gamma_{t\varphi}^2 \right)^{-1/4}.
	\label{EqOmegaRescaling}
\end{align}
Under this rescaling, the Wheeler DeWitt equation becomes
\begin{align}
	\left( \hbar^2 \left( \frac{1}{4} \frac{\partial^2}{\partial \gamma_{t\varphi^2}} - \frac{\partial}{\partial \gamma_{tt}} \frac{\partial}{\partial \gamma_{\varphi \varphi}} + \frac{1}{2 \gamma_{\varphi \varphi}} \frac{\partial}{\partial \gamma_{tt}} \right) + \frac{1}{4\kappa^2 l^2} \right) \Psi = 0,
	\label{EqWdWRescaled}
\end{align}
and the norm associated with $\gamma_{\varphi \varphi}$ as a clock
\begin{align}
	|\Psi|^2_{\gamma_{\varphi\varphi}} =\frac{i}{2} \int d\gamma_{tt} d\gamma_{t\varphi} \frac{1}{\gamma_{\varphi \varphi}} \left( \Psi^* \partial_{\gamma_{tt}} \Psi - \text{h.c.} \right).
	\label{EqRescalednorm}
\end{align}
The rescaling by $\Omega$ as in \eqref{EqOmegaRescaling} has another upshot. We remarked earlier that our basis of solutions \eqref{EqPsiBasis}  only solve \eqref{EqWDWequation} 
to leading semiclassical order. However, we note that the basis \eqref{EqPsiBasis} solves the rescaled equation \eqref{EqWdWRescaled} to all orders in $\hbar$. That is to say, these states were only really an \textit{approximate} basis of solutions for \eqref{EqWDWequation}, where we had ignored subleading pre-factors. Those pre-factors are in fact exactly absorbed by the lapse rescaling \eqref{EqOmegaRescaling}, so that \eqref{EqPsiBasis} is an \textit{exact} basis for \eqref{EqWdWRescaled}. So the lapse rescaling not only simplifies computations, but is also the most convenient choice for keeping track explicitly of the $\hbar$ dependence of our quantum corrections. 

For a general solution \eqref{EqPsiGeneralcJ}, we evaluate the norm \eqref{EqRescalednorm} to be (from here onwards dropping the explicit label to the $\gamma_{\varphi\varphi}$ clock) 
\begin{align}
	\left| \Psi \right|^2 = 2\kappa \hbar \int \frac{dc \, dj}{(2\pi)^2} c \left| \beta(c,j) \right|^2.
	\label{EqNormBeta}
\end{align}
To study the quantum spread of the wavefunction, we will in particular study the metric function $\gamma_{tt}$, its variance and its conjugate momentum $p^{tt}$. 
It will be useful to introduce a slightly redefined wavepacket $\beta$, 
\begin{align}
	\begin{split}
		\beta(c,j) = & N_{\beta} \sqrt{c} \exp \left \lbrace - i m_0 (c-c_0) - \frac{\Delta_c^2}{2} \left( c- c_0 \right)^2 \right \rbrace \\
		& \times \exp \left \lbrace - i w_0 j - \frac{\Delta_j^2}{2} (j-j_0)^2 \right \rbrace,
	\end{split}
	\label{EqBetaDef2}
\end{align}
with $N_{\beta} = 1/\sqrt{\pi \kappa \hbar \left( 1 + 2 c_0^2 \Delta_c^2 \right)/(\Delta_c^3 \Delta_j)}$. 
This wavepacket is strongly peaked on $c_0$ and $w_0$ for $\Delta_c \gg 1$ and $\Delta_j \ll 1$.  
The leading order semiclassical behaviour of this wavepacket is thus still \eqref{EqOnShellWavefunction}; we have additionally now specified the next-to-leading order behaviour such that when \eqref{EqNormBeta} is evaluated on \eqref{EqBetaDef2}, the norm is $\left| \Psi \right|^2 = 1$. 

We can now proceed to evaluate the expectation value of the metric function $\gamma_{tt}$. In general, we have (now reinstating $\epsilon = \pm 1$) 
\begin{align}
	\begin{split}
		\langle \gamma_{tt} \rangle =  \epsilon \, 4 \kappa^2 \hbar \int \frac{dc dj}{(2\pi)^2} & \left( 2 c \gamma_{\phi \phi} \partial_j \beta(c,j) \partial_j \beta^*(c,j) - \frac{2c^3 \gamma_{\phi \phi}}{l^2} \left| \beta(c,j) \right|^2 \right. \\
		& \left. + i \left[ c^2 \hbar \beta(c,j) \left( j \partial_j \beta^*(c,j) - c \partial_c \beta^*(c,j) \right) - \text{h.c.} \right] \right),  \label{gttvevint}
	\end{split}
\end{align}
which on our new wavepacket evaluates to
\begin{align}
	\begin{split}
		\langle \gamma_{tt} \rangle &=  \epsilon \, 4 \kappa^2 \left( \gamma_{\varphi \varphi} \left( w_0^2 - \frac{c_0^2}{l^2} \right) + c_0 \left( c_0 m_0 - j_0 w_0 \right) \right) + \mathcal{O} \left( \hbar^2, \frac{1}{\Delta_c^2} , \Delta_j^2 \right) \\
		&=  \left. \gamma_{tt} \right|_{\text{classical}}  + \mathcal{O} \left( \hbar^2, \frac{1}{\Delta_c^2} , \Delta_j^2 \right).  
	\end{split}    \label{gttvevresult}
\end{align}
We may similarly evaluate the variance
\begin{align}
	\langle \gamma_{tt}^2 \rangle - \langle \gamma_{tt} \rangle^2 = 8c_0^2 \kappa^4 \left( \frac{w_0^2}{\Delta_j^2} + \hbar^2 \Delta_c^2 \right) + \mathcal{O} \left( \hbar^2, \frac{1}{\Delta_c^2}, \Delta_j^2 \right).
	\label{Eqgttvar}
\end{align}
Near the horizon ($\gamma_{tt}|_{\text{classical}} \rightarrow 0$), we find that 
\begin{align}
	\left. \frac{\langle \gamma_{tt}^2 \rangle - \langle \gamma_{tt} \rangle^2}{\langle \gamma_{tt} \rangle^2} \right|_{\langle \gamma_{tt} \rangle \rightarrow 0} = \frac{2 c_0^2\left( c_0^2 - l^2 w_0^2 \right)^2}{l^4 \left( c_0 m_0 - j_0 w_0 \right)^2 \hbar^2} \frac{\Delta_c^2}{\Delta_j^4} + \mathcal{O} \left( \hbar^0, \Delta_j^0, \Delta_c^0 \right).
	\label{Eqgttvar2norm}
\end{align}
Recall that $\hbar, \Delta_j \ll 1$ and $\Delta_c \gg 1$. Thus, the significance of \eqref{Eqgttvar2norm} is that the variance is significantly larger than the expectation value of $\langle \gamma_{tt} \rangle$ near the horizon. That is, the spread of the wavefunction becomes 
very large around the horizon, and quantum effects dominate over semiclassical ones. 

We verify the result \eqref{Eqgttvar} by computing the expectation value of its conjugate momentum and recovering the Heisenberg uncertainty limit. As noted in earlier discussion, there is an ordering ambiguity when defining momentum operators such as $p^{tt} = - i \hbar \partial_{\gamma_{tt}}$. We choose a prescription consistent with our Laplacian normal ordering that recovers a real answer
\begin{align}
	\begin{split}
		\langle p^{tt} \rangle &= \left( - i \hbar \right) \frac{i}{2} \int d\gamma_{tt} d\gamma_{t\varphi} \frac{1}{\gamma_{\varphi \varphi}} \left( \Psi^* \partial_{\gamma_{tt}} \left( \partial_{\gamma_{tt}} \Psi \right) - \partial_{\gamma_{tt}} \Psi \partial_{\gamma_{tt}} \Psi^* \right) \\
		&= - \frac{\epsilon \hbar}{2\kappa} \int  \frac{dc dj}{(2\pi)^2} \left| \beta(c,j) \right|^2    \label{pttvevint}
	\end{split}
\end{align}
which gives 
\ali{
	\langle p^{tt} \rangle &= -\epsilon \frac{1}{4 \kappa^2 c_0} + \mathcal O(\Delta_c^2) = \left. p^{tt} \right|_{\text{classical}} + \mathcal O(\Delta_c^2). \label{pttcalc}
}
We similarly compute the variance to be
\begin{align}
	\langle (p^{tt})^2 \rangle - \langle p^{tt} \rangle^2 &= \frac{1}{32 c_0^4 \kappa^4 \Delta_c^2} + \mathcal{O} \left( \frac{1}{\Delta_c^3} \right) 
\end{align}
and the uncertainty product 
\begin{align}
	\left( \langle (p^{tt})^2 \rangle - \langle p^{tt} \rangle^2 \right) \left( \langle \gamma_{tt}^2 \rangle - \langle \gamma_{tt} \rangle^2 \right) = \frac{\hbar^2}{4} + \mathcal{O} \left( \frac{1}{\Delta_c^2}, \Delta_j^2 \right) . 
\end{align}
We thus have a state which saturates the uncertainty principle, which is to be expected for a Gaussian wavepacket. This computation verifies our computation of the variance in $\langle \gamma_{tt} \rangle$.

\section{Energy discussion: CFT and $T\bar T$ interpretations}   \label{sectenergy} 

In Sections \ref{sect3} and \ref{sectsc}, we have constructed a WdW state for the BTZ geometry, following the program of \cite{Hartnoll:2022snh,Blacker:2023oan,BlackerNing:2023ezy}.  As in those works, we expect this state to have a dual interpretation as a QFT path integral. In \cite{Hartnoll:2022snh,Blacker:2023oan,BlackerNing:2023ezy}, it was conjectured what the holographic relationship was, and evidence was gathered about the properties of the dual  theory. Here, in a $(2+1)$-dimensional set-up, we  can instead rely on earlier literature \cite{McGough:2016lol} to identify the dual theory more precisely with a $T\bar T$-deformed theory. In this section, we calculate gravitational energy measures for the BTZ solution of Section \ref{sect3}, including the equivalent of the energy measure discussed in \cite{Hartnoll:2022snh}. We  discuss their holographic interpretations in terms of a CFT dual at the conformal boundary and a $T\bar T$ dual on the $\Sigma$ slice. In particular, we review the match to the $T\bar T$ spectrum, first presented in \cite{McGough:2016lol}.

\begin{figure}[t]
	\centering 
	\includegraphics[width=9cm]{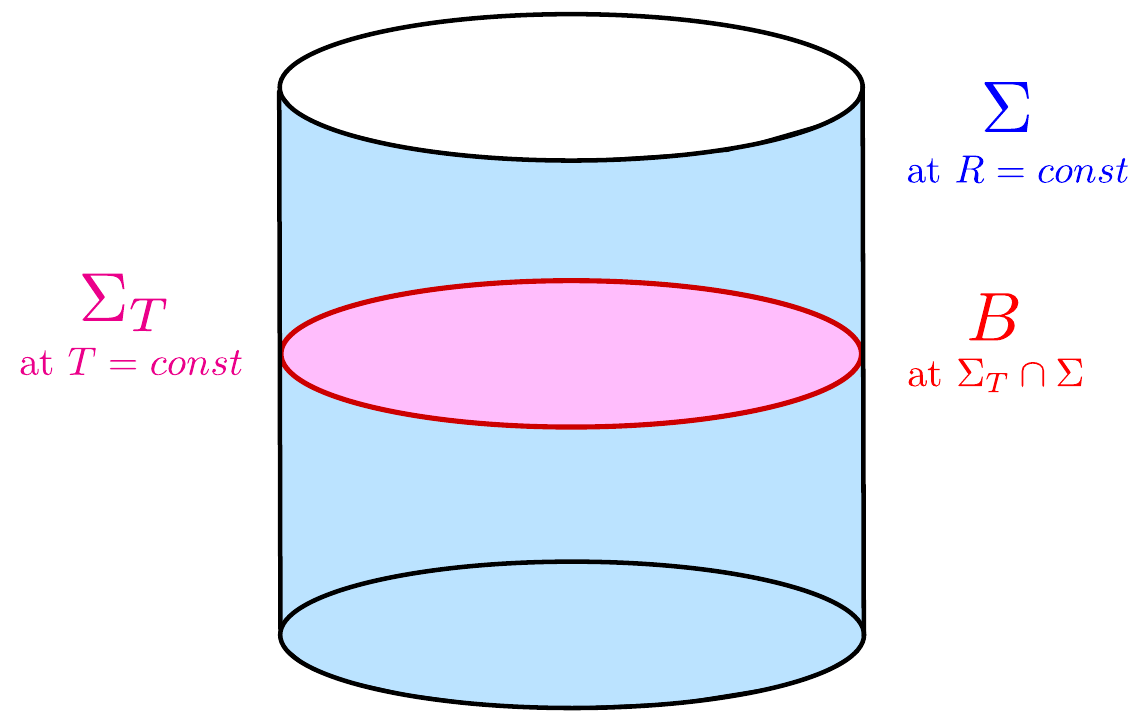} 
	\caption{Visualization of different surfaces that enter the energy discussion: 2-dimensional $\Sigma$ (blue) at constant $R$, 2-dimensional $\Sigma_T$ (pink) at constant $T$ and 1-dimensional $B$ (red) at their intersection. 
	} 
	\label{figSigmasurfaces} 
\end{figure}

\paragraph{Brown-York stress tensor, conserved charges and energy} 
The Brown-York stress tensor \eqref{BYdef} for a solution is given by 
\ali{
	T^{ij} = \frac{2}{\sqrt{-\gamma}} \pi^{ij},  
}   
with the contribution from the counterterm $S_{ct}$ included by using \eqref{cantranf}. 
It measures the gravitational energy-momentum of the solution in the region bounded by the 
hypersurface $\Sigma$ with induced metric $\gamma_{ij}$. We will in this section use both the $(r,t,\varphi)$ coordinates and $(R,T,\phi)$ coordinates of Section \ref{sect3}, and write out  when necessary the components of $\gamma_{ij}$ and $T^{ij}$ to make clear which of the coordinates are being used. We consider $\Sigma$ to be a hypersurface in the exterior region of the black hole at constant $r$, or equivalently, at constant $R$. 
The induced metric on a constant-$R$ hypersurface $\Sigma$ of the metric \eqref{BTZsol} is $\gamma_\mn = g_\mn - n_\mu n_\nu$ in terms of the normal $n_\mu = \delta_\mu^R / \sqrt{f_{m,j}(R)}$.

Following the volume time discussion in Section \ref{sect2}, 
let us also introduce the trace-free part  
$\tilde T^{ij}$ of the Brown-York tensor through 
\ali{
	& T_{ij} = \tilde T_{ij} + \frac{1}{d} \gamma_{ij} \Theta .  
}
It is the stress tensor associated with the metric $\tilde \gamma_{ij}$,  
\ali{
	\tilde T^{ij} = \frac{2}{\sqrt{-\tilde \gamma}} \tilde \pi^{ij} = 2 \, \tilde \pi^{ij},   
}
with $\tilde \gamma_{ij}$ and $\tilde \pi^{ij}$ defined in \eqref{gammatilde}-\eqref{orthdec}. 
All objects with tildes are tensors whose indices are raised and lowered with the determinant-one metric $\tilde \gamma_{ij}$. Therefore, $\tilde T^i_j = \sqrt{-\gamma}\, T^i_j - \frac{1}{d} \sqrt{-\gamma}\, \delta^i_j \Theta $.

With the Brown-York tensor, one can define an energy density $\mathcal E$, with respect to the time $T$ of the BTZ solution. The definition  
naturally requires the introduction of constant-$T$ slices $\Sigma_T$ with induced metric $h_{ab}$ (with $a,b = R,\phi$). 
In an ADM decomposition notation, we have $ds^2 = -N_T^2(R) dT^2 + h_{ab}(R) dx^a dx^b$, 
with $N_T^2 \equiv f_{m,j}$ and $h^\phi_T = -j/(2R^2)$, as can be read off from \eqref{BTZsol}. The normal to $\Sigma_T$ is given by 
$u_\mu = -N_T \delta_\mu^T$, in terms of which the induced metric is $h_\mn = g_\mn + u_\mu u_\nu$. 
The proper energy density $\mathcal E$ is then given by the projection  \cite{Brown:1992br,Brown:1994gs} 
\ali{
	\mathcal E = u_i u_j T^{ij} 
	= u_T^2 T^{TT} = N_T^2 T^{TT} = f_{m,j} T^{TT} .   \label{epsilonformula}
}
The corresponding energy for $\Sigma$ is 
\ali{
	E = \int_B d^{D-2} x \sqrt{\sigma} u_i u_j T^{ij} \label{Eformula}
}
with $\sigma_\mn = \gamma_\mn + u_\mu u_\nu$ the intrinsic metric of the constant-$R$, constant-$T$, (D-2)=1-dimensional hypersurface $B$ at the intersection of $\Sigma$ and $\Sigma_T$. The different hypersurfaces $\Sigma$, $\Sigma_T$ and $B$ are shown in Figure \ref{figSigmasurfaces}. 

The conserved charge associated with a Killing vector field $\xi$ on $\Sigma$ is \cite{Brown:1992br,Brown:1994gs} 
\ali{
	Q_\xi = \int_B d^{D-2} x \sqrt{\sigma} \xi_i u_j T^{ij}.  \label{Qdef}
}
For the Killing vector $\xi_{(T)} \equiv \p_T$, it gives the conserved BTZ mass $M \equiv Q_{\xi_{(T)}}$, and for $\xi_{(\phi)} \equiv \p_\phi$, the conserved BTZ angular momentum $J \equiv -Q_{\xi_{(\phi)}}$.  
Using $u \cdot \xi_{(T)} = -N_T$, the mass can be written as 
\ali{
	M = - \int d\phi \sqrt{-\gamma} \, T^T_T  , \label{BTZM}
}
the angular momentum as 
\ali{
	J = \int d\phi \sqrt{-\gamma} \, T^T_\phi  , \label{BTZJ}
}
and the energy as 
\ali{
	E = \int d\phi \sqrt{-\gamma} N_T T^{TT}  \label{energyE}
}
with $\gamma = \gamma_{TT}\gamma_{\phi\phi} - \gamma_{T\phi}^2$.  
In the non-rotating case, the mass and energy differ only by a lapse factor,  $M = N_T E$, and in the rotating case by 
\ali{
	E = \frac{1}{N_T} (M - \frac{j}{2R^2} J) . \label{energy} 
} 

In the asymptotic limit $R\ra \infty$ where $\Sigma$ reaches the boundary, the conserved mass is given by the constant ADM mass 
\ali{
	M|_\infty = \frac{\pi}{\kappa} m . \label{mADM} 
}
Similarly, the constant $j$ in the BTZ solution \eqref{BTZsol} is the conserved charge associated with asymptotic invariance under rotations 
$J|_\infty = \pi j/\kappa$. 
That is, the 
constants 
$m$ and $j$ of the BTZ solution are the conserved charges associated with asymptotic invariance under time translations and rotations, i.e.~mass and angular momentum \cite{Banados:1992wn}.

\subsection{Dual CFT energy }  
We now turn to a discussion of the dual CFT energy, which starts with an application of the strategy used in \cite{Hartnoll:2022snh} to our rotating solution. 
The on-shell action $I_{cl}$ as a function of the boundary data is equal to the Hamilton's principal function $\mathcal I$ in \eqref{EqS}, with $\epsilon = -1$ in the exterior  \cite{Papadimitriou:2016yit}. The effect of the counterterm can be included by performing the canonical transformation \eqref{cantranf}, which amounts to 
an on-shell action $S_{cl} = I_{cl} + S_{ct}$ equal to the Hamilton's principal function $\mathcal S$ for the momenta $\pi^{ij} = \partial_{\gamma_{ij}} \mathcal S(\gamma_{ij})$,  
\ali{
	S_{cl}[\gamma_{tt},\gamma_{\varphi\varphi}, \gamma_{t\varphi}; c,j] = \mathcal S ,  
}
with
\begin{align}
	&\mathcal S \left(\gamma_{ij};c,j \right) = \frac{1}{4 \kappa^2 c} \gamma_{tt} + \frac{c}{l^2} \gamma_{\varphi \varphi} + \frac{\left( c j - \gamma_{t\varphi}/\kappa \right)^2}{4c \gamma_{\varphi \varphi}} - \frac{\sqrt{-\gamma_{t\varphi}^2 - \gamma_{tt}\gamma_{\varphi\varphi}}}{l \kappa} . 
	\label{EqSinclct}
\end{align}
This object has the invariance  
\ali{
	\mathcal S[\gamma_{tt} e^{-k},\gamma_{\varphi\varphi} e^{k}, \gamma_{t\varphi}; c \,  e^{-k},j  e^{k}] = 	\mathcal S[\gamma_{tt},\gamma_{\varphi\varphi} , \gamma_{t\varphi}; c,j] . 
}
From this, it follows that 
$\p_k \mathcal S|_{k=0} = 0$,  which can be written as
\ali{
	- c \, \p_{c} \mathcal S + j \,\p_j \mathcal S = \gamma_{tt} \frac{\delta \mathcal S}{\delta \gamma_{tt}} - \gamma_{\varphi\varphi} \frac{\delta \mathcal S}{\delta \gamma_{\varphi\varphi}} 
} 
or still as 
\ali{
	- c \, \p_{c} \mathcal S  
	+ j \p_j \mathcal S = \frac{\sqrt{-\gamma}}{2} (T^t_t - T^\varphi_\varphi)  .  
}
The first term on the left hand side can also be written as $\p_{k_0} \mathcal S$, with the constant $c = -e^{-k_0}$, to connect to the notation used in \cite{Hartnoll:2022snh}. The right hand side is equal to $\sqrt{-\gamma}  (T^t_t - \frac{\Theta}{2})$, 
or in terms of the  traceless tensor $\tilde T_{ij}$, 
\ali{
	- c \, \p_{c} \mathcal S  
	+ j \p_j \mathcal S = \tilde T^t_t. \label{expr}   
}
By HJ construction, the expression is constant. More precisely, from \eqref{HJcsts},  
it is equal to $- c\, m + j w$. On the right hand side, the constant can be expressed either as $\tilde T^t_t$ or as the asymptotic limit $\lim_{r \ra \infty}  \sqrt{-\gamma}  \, T^t_t$, using that 
in that limit,  
the trace of the stress tensor of the solution vanishes, 
$\lim_{r \ra \infty} \Theta = 0$ (for our flat slicing). 

Through the holographic dictionary, the Brown-York stress tensor for the asymptotic 
radial slice $\Sigma$ 
calculates the expectation value of the dual CFT stress tensor. As such, the integrated right hand side of \eqref{expr} calculates the CFT charge 
\ali{
	\vev{Q}^{CFT}_{\xi_{(t)}} = \int d\varphi \, \vev{\tilde {T}^t_t}_{CFT}
} 
associated with 
translations in $t$, to be given by $2 \pi (-c m + w j)$. 

Let us also pause explicitly at the interpretation in terms of the BTZ coordinates $(R,T,\phi)$, used in \eqref{BTZsol}. 
From the coordinate transformation \eqref{EqPhiTtransform}, we have 
\ali{
	T^t_t = T^T_T + \frac{w}{c} T^T_\phi, \qquad T^\varphi_\varphi = T^\phi_\phi - \frac{w}{c} T^T_\phi
}
such that 
\ali{
	2\pi 
	\left( - c \, \p_{c} \mathcal S + j \p_j \mathcal S \right) & = 2 c \kappa \int d\phi \sqrt{-\gamma} \left(T^T_T + \frac{w}{c} T^T_\phi - \frac{1}{2} \Theta \right) \label{firstline} \\
	&= 2 c \kappa \int d\phi \left(\tilde T^T_T + \frac{w}{c} \tilde T^T_\phi \right). \label{secondline}
}
The factor of $2 c \kappa$ comes from a factor $dT/dt$ in the measure, as $\gamma$ in these lines now 
refers to the determinant in the $(T,\phi)$ coordinates. Using the definitions \eqref{BTZM} and \eqref{BTZJ}, 
the first line \eqref{firstline} becomes 
\ali{
	2\pi 
	\left( - c \, \p_{c} \mathcal S + j \p_j \mathcal S \right) & = 2 c \kappa \left(-M + \frac{w}{c} J - \frac{1}{2} \Theta \right) . 
}
All terms on the right hand side are $R$-dependent, $M \equiv M(R)$, etc.  
But again, using $\lim_{r \ra \infty} \Theta = 0$ for our slicing, the constant expression is equal to $2 c \kappa (-M|_\infty + \frac{w}{c} J|_\infty)$, or by \eqref{mADM} for the ADM mass, indeed equal to $2 \pi (-c m + w j)$ on the left hand side. This is a consistency check on the definitions of $m$ and $j$ from the HJ perspective in \eqref{HJcsts} on one hand, and as ADM mass \eqref{mADM} and angular momentum from the Noether charges in \eqref{Qdef}-\eqref{BTZJ} on the other. 

From \eqref{secondline}, it follows that the ADM mass $\frac{\pi}{\kappa} m$  
and momentum $\frac{\pi}{\kappa} j$ respectively calculate 
the classical values of the CFT charges ($Q_\xi^{CFT} = \int d\phi \,  j^T_{CFT}$)  associated with translations in time $T$ and angular direction $\phi$,    
\ali{
	\vev{Q}^{CFT}_{\xi_{(T)}} = -\int d\phi \, \vev{\tilde {T}^T_T}_{CFT}, \qquad 
	\vev{Q}^{CFT}_{\xi_{(\phi)}} = \int d\phi \, \vev{\tilde {T}^T_\phi}_{CFT} ,  
}
for conformal currents $j^i_{CFT} = \tilde T_{CFT}^{ij} \xi_j$.

\subsection{Dual $T\bar T$ energy }  
Now we turn to the energy discussion for non-asymptotic radial slices $\Sigma$ of the BTZ metric \eqref{BTZsol}, at a constant value of $R$. 
The induced metric on $\Sigma$ is $ds^2_\Sigma = - f_{m,j}(R) dT^2 + R^2 \left( d{\phi} - \frac{j}{2R^2} d{T} \right)^2$, which can be written in terms of new coordinates $(\tilde t, \tilde \varphi)$ as 
\ali{
	ds^2_\Sigma = -d\tilde t^2 + R^2 \, d\tilde \varphi^2   \label{ds2Sigma}
}
with 
\ali{
	\tilde t = N_T T, \qquad \tilde \varphi = \phi - \frac{j}{2R^2} T . 
}
By the cut-off holography proposal of \cite{McGough:2016lol}, there is a dual interpretation of the BTZ solution with boundary $\Sigma$ at a \emph{finite} value of $R$ in terms of a $T\bar T$ theory living on the fixed boundary metric $ds^2_\Sigma$, with $R$ in the role of the radius of the cylindrical geometry. One of the main arguments for this proposal is the identification of the cut-off BTZ energy spectrum with the $T\bar T$ spectrum. We repeat this argument here, focusing on the difference between energy $E$ and mass $M$.

For the $T\bar T$ spectrum, we look at the $T\bar T$ conserved charge associated with translations in $\tilde t$,  
\ali{
	\vev{Q}^{QFT}_{\xi_{(\tilde t)}} = \int d\tilde \varphi \, \sqrt{\sigma} \, \vev{ {T}^{\tilde t}_{\tilde t}}_{QFT}  \label{QQFT}
}  
with 
$\sigma =R^2$ the determinant of the metric $ds^2_B = R^2 \, d\tilde \varphi^2$ on the constant-time hypersurface $B$. 
On the bulk side, this is calculated by the conserved charge $Q_{\xi_{(\tilde t)}}$ for the Killing vector $\xi_{(\tilde t)} = \p_{\tilde t}$, or mass 
\ali{
	M_{\tilde t} \equiv Q_{\xi_{(\tilde t)}} = \int d\tilde \varphi \sqrt{\sigma} u_{\tilde t} T^{\tilde t}_{\tilde t}   
} 
where we used the definition \eqref{Qdef}. 
To distinguish it explicitly from the mass charge $Q_{\xi_{(T)}}$ introduced in \eqref{BTZM}, let us in this section use the notation $M_T$ for \eqref{BTZM} and $J_\phi$ for \eqref{BTZJ}.  Then we find by simple tensor calculus the following relation\footnote{For $t' = a t$, $\theta' = c \theta + b t$, we have $M_{t'} = \frac{1}{a} (M_t + \frac{b}{c} J_\theta)$.} between $M_{\tilde t}$, $M_T$ and $J_\phi$:   
\ali{
	M_{\tilde t} = \frac{1}{N_T} (M_T - \frac{j}{2R^2} J_\phi).  
}
This can be recognized as the energy of the constant-$R$ $\Sigma$ slices with respect to bulk time $T$, given as $E$ in \eqref{energy}, 
\ali{
	M_{\tilde t} = E.  \label{massttilde}
} 
This is the reason it is the energy formula \eqref{Eformula} that is used (in \cite{McGough:2016lol,Marolf18,Jiang:2019epa} etc.) to calculate what can be interpreted as the $T\bar T$ spectrum. We are just making very explicit here the comment in \cite{McGough:2016lol} that $E$ calculates the mass charge for  $ds^2_\Sigma$ with lapse equal to one (in \eqref{ds2Sigma}). 

Using the formulas \eqref{epsilonformula} for the energy density and \eqref{Eformula} for $E$, we recover 
\ali{
	& \mathcal E = -\frac{\sqrt{f_{m,j}(R)}}{\kappa R} + \frac{1}{\kappa l}   
}
and \ali{ 
	& E = \int d\phi R \, \mathcal E = -\frac{2\pi \sqrt{f_{m,j}(R)}}{\kappa} + \frac{2\pi R}{\kappa l}, 
	\label{ETTbarmatch}
}
with the function $f_{m,j}$ given in \eqref{fmj}. 
The second term in these expressions is the contribution coming from the counterterm in the action. It is such that in the limit of $R \ra \infty$, both $\mathcal E$ and $E$ vanish, as $\sqrt{f_{m,j}(R)} \ra R/l$. It is the dimensionless object $E R$ that asymptotes to $l M|_\infty$. 
The $T\bar T$ spectrum $E_{QFT}$ on a cylinder of size $L$ (equal to $R$ in \eqref{ds2Sigma}) 
is known as a function of the $T\bar T$ coupling $\lambda$ and 
the seed CFT's energy and momentum, given holographically by $M|_\infty$ in \eqref{mADM} and $J|_\infty$ below equation \eqref{mADM}. 
In \cite{Marolf18,McGough:2016lol}, the dimensionless combination $E R$ is matched to the $T\bar T$ spectrum $E_{QFT} L$ for the $\lambda$ identification \eqref{lambdaVMM}, and we find agreement with that conclusion.

In this section we have focused on the BTZ solution, and identified the bulk mass or conserved charge \eqref{massttilde} that calculates the $T\bar T$ energy spectrum \eqref{ETTbarmatch} for $T\bar T$ coupling $\lambda = -\frac{1}{2} \kappa l$. 
This constitutes a traditional holographic match between conserved charges on each side of the duality (which is slightly obscured by using the energy formula \eqref{Eformula} rather than \eqref{Qdef} in e.g.~\cite{McGough:2016lol,Jiang:2019epa}). 
It adds an explicit entry 
\ali{
	M_{\tilde t} 
	= E_{QFT} 
}  
to the duality $\psi = Z_{QFT}$ in \eqref{ZQFT} applied to the BTZ wavefunction \eqref{EqOnShellWavefunction}, with `QFT' referring to the dual $T\bar T$ theory.

\subsection{WdW perspective on $T\bar T$ energy levels} 
It is well-known that due to the square root structure of the $T\bar T$ energy levels, 
these can become imaginary. The standard cut-off holographic interpretation of this is the following: the gravitational energy levels \eqref{ETTbarmatch} of cut-off BTZ become imaginary when the cut-off crosses the BTZ horizon and the BTZ solution as such no longer `fits' in the imposed holographic box. 
From the WdW perspective, one can argue that the imaginary energy levels are an allowed feature of the theory, and correspondingly that the holographic interpretation of the WdW wavefunction as a $T\bar T$ partition function holds even when the bulk cut-off crosses the horizon.   
Let us expand on this argument. 

In the discussion of the semi-classical states for BTZ, we chose in Section \ref{SecSemiclassicalStates} to construct the \emph{Vilenkin} wavefunction $\Psi$ given in \eqref{EqPsiBasis2}. This is the wavefunction with $(+i \mathcal I)$ in the exponent. It describes the set of hypersurfaces $\Sigma$ that cover the blue and red region in Fig.~\ref{figBTZnonrot}, i.e.~that cover the outside-horizon region and the future black hole interior. Thinking of these hypersurfaces as evolving in the $r$-direction, the Vilenkin wavefunction \eqref{EqPsiBasis2} corresponds to `upwards' evolution along $r$ in Fig.~\ref{figBTZnonrot}. The other WKB solution, 
with $(-i \mathcal I)$ in the exponent, similarly would correspond to `downwards' evolution along $r$ in Figure \ref{figBTZnonrot}, associated with the set of hypersurfaces $\Sigma$ covering the outside-horizon and the past black hole interior.  
Since $r$ becomes a timelike direction past the horizon, the choice of sign in the exponent of our WKB solution \eqref{EqPsiBasis2} corresponds to a breaking of time symmetry at the crossing of the horizon. 
This leads to a CPT symmetry breaking and associated non-unitarity, which in turn is reflected in energy levels becoming complex. This is entirely consistent with the $T\bar T$ interpretation, as the dual $T\bar T$ energy levels become complex precisely at the point of crossing the horizon in the bulk description. 
This provides the WdW wavefunction point of view on the origin of imaginary energy levels in $T\bar T$. In \cite{WallAraujo-Regado:2022gvw}, such a point of view was discussed for Cauchy slices, whereas our discussion here is for the radial slicing in Fig. \ref{figBTZnonrot} that we consider in this paper. 

Entirely analogous, the red regions in Fig.~\ref{figBTZrot} where the $r$ coordinate of the \emph{rotating} BTZ solution becomes timelike correspond to imaginary $T\bar T$ energy levels in the dual description ($f_{m,j} < 0$).

The preceding argument makes use of the classical solution $\mathcal I$ in \eqref{EqS}, exponentiated to give a WKB solution \eqref{EqPsiBasis2}. More precisely, \eqref{EqPsiBasis2} is a Vilenkin wavefunction solution $\Psi$ to the WdW equation to leading semi-classical order. In Section \ref{sectsc}, we included some quantum effects by discussing  
the construction of more general wavepacket solutions \eqref{EqPsiGeneralcJ} that are peaked around the classical solution. We will here investigate if the wavepacket solutions provide additional information about the $T\bar T$ spectrum, particularly as the horizon is crossed.

We constructed a particular wavepacket solution with large $\Delta_c$ and small $\Delta_j$ (centered on classical values $c_0$ and $w_0$ of $c$ and $w$) by using $\beta$ defined 
in \eqref{EqBetaDef2}. It is such that the DeWitt norm of the wavepacket solution $\Psi$ 
in \eqref{EqNormBeta} is one. This norm is defined with respect to a particular choice of clock or `time' in mini-superspace, namely $\gamma_{\phi\phi}$. This is not only a natural choice of clock from the bulk perspective, as explained in Section \ref{sect43}, but also from the dual $T\bar T$ perspective, as we discuss next.

The $T\bar T$ coupling $\lambda$ given in \eqref{lambdaVMM} has dimensions length squared. For $T\bar T$ on a cylinder  dual to cut-off BTZ gravity, the dimensionless combination that appears in the $T\bar T$ energy expression is the $T\bar T$ coupling divided by the cylinder radius squared. From the dual bulk perspective, this is the dimensionless combination 
\ali{
	\hat \lambda = \frac{\lambda}{R^2} 
}
with $R$ the radial coordinate of BTZ \eqref{BTZsol}. This makes it particularly clear that the $T\bar T$ deformation corresponds to moving inwards into the holographic bulk. It also makes that the choice of clock $\gamma_{\varphi\varphi} \equiv R^2$  
in \eqref{EqRescalednorm} can be recognized as describing $T\bar T$ evolution  $\hat \lambda \sim 1/\gamma_{\varphi\varphi}$. 
Moreover, it is the more well-behaved choice of clock than the volume time $r = \pm v l$ in \eqref{volumetimeBTZ} to investigate near-horizon physics, since volume time switches sign at the horizons as pictured in Figure \ref{figvolBTZ}. This volume time was argued (from an ADM reparametrization perspective) to provide the natural bulk evolution dual to $T\bar T$ in the \emph{semi-classical} regime. In this regime, the WdW equation for the $T\bar T$ partition function can be rewritten 
as a Schr\"odinger equation in volume time. Near the boundary of the BTZ solution, its volume time and $\gamma_{\varphi\varphi}$ time indeed coincide, as they should: $v \sim R^2$ in the limit $R \ra \infty$. Volume time is the useful choice of time to describe the bulk evolution corresponding to the initial $T\bar T$ deformation of a holographic CFT, but to probe the bulk deeper we use $\gamma_{\varphi\varphi} \equiv R^2$ (see also the discussion in the last paragraph of Section \ref{sect2TTbar}). 

Now that we have established that $\gamma_{\varphi\varphi}$ is a sensible choice of time for $T\bar T$ interpretations, we revisit the results of Section \ref{sect43} from this perspective. 

The gravitational energy $E$ in \eqref{energyE} that calculates the dual $T\bar T$ energy can be written in terms of the momenta as 
\ali{
	E \sim \sqrt{\gamma_{tt}} \, \pi^{tt}. 
	}
On the wavepacket, the expectation value of the energy in the $\gamma_{\varphi \varphi}$ clock is $\vev{E} \sim \vev{\sqrt{\gamma_{tt}}  \pi^{tt}}$. Here, $\pi^{tt}$ differs by the contribution of the counterterm from $p^{tt}$, the expectation value of which was calculated in \eqref{pttcalc}. In terms of the coordinates $\gamma_{ij}$ and momenta $p^{ij}$, there are thus contributions to the energy of the form  $\vev{\sqrt{\gamma_{tt}} p^{tt}}$ and of the form $\vev{\sqrt{-\gamma_{tt} \gamma}  \gamma^{tt}}$. The notation is short for expectation values with respect to `time' $\gamma_{\varphi \varphi}$, i.e.~using the norm \eqref{EqRescalednorm}. Both of these contributions can in principle be calculated using the techniques of Section \ref{sect43}, there put to use to obtain $\vev{\gamma_{tt}}$ and $\vev{p^{tt}}$. In practice, however, the equivalent expressions to \eqref{gttvevint} and \eqref{pttvevint} are very hard to obtain, and we have not succeeded in directly calculating $\vev{E}$ for the wavepacket as a result. That said, in \eqref{Eqgttvar2norm} we obtained the result that the variance of $\gamma_{tt}$ becomes so large around the horizon that quantum effects dominate. It is highly likely that the more complicated 
terms contributing to $E$, both dependent on $\gamma_{tt}$, will similarly have a variance that blows up around the horizon. Indeed, it would be sufficient for either the $p^{tt}$ contribution \emph{or} the counterterm contribution to $E$ to have such a near-horizon behavior to cause a breakdown of the wavepacket description. Note however that the counterterm contribution \emph{classically} only is responsible for the addition of a term $2 \pi R/(\kappa l)$ to the energy in \eqref{ETTbarmatch}, and as such plays no role in the classical discussion of imaginary energy levels due to the square root structure $\sqrt{f_{m,j}(R)}$.

To summarize, the energy levels becoming complex beyond the horizon can be understood in terms of a CPT breaking argument at the classical level. On the wavepacket \eqref{EqBetaDef2} centered on the classical solution, the result \eqref{Eqgttvar2norm} for the expectation value and variance of $\gamma_{tt}$ indicates that the wavepacket description of the energy levels is highly likely to break down near the horizon, where quantum effects will instead dominate. \\

\section{Discussion and outlook} 

In this work, we have studied the radial canonical formalism of asymptotically AdS$_3$ gravity. In its original form, the WdW equation is a relation between metric functions on a slice of spacetime, and with no explicit time parameter is in a sense ``timeless''. By employing a deparametrization strategy, as in \cite{Arnowitt:1962hi}, we showed in Section \ref{sect2} that in a radial slicing the volume density of the radial slices can in fact be interpreted as a time, and the WdW equation 
as a Schr\"odinger equation. From a holographic point of view, we concluded that at the semi-classical level, the Lorentzian partition function of $T \bar T$ theory satisfies a Schr\"odinger equation in volume time describing evolution into the bulk, and the $T \bar T$ operator expectation value can be naturally interpreted in terms of the bulk volume time Hamiltonian.

We looked to test our insight from our volume time interpretation by studying WdW states of a (2+1)-dimensional theory of gravity and their dual $T\bar T$ interpretation. In particular, in Sections \ref{sect3} and \ref{sectsc} we turned our attention to the BTZ solution, constructing WdW states from a Hamilton-Jacobi function. A technical upshot of this computation is that we were able to include rotation of the black hole solution. We constructed both the classical solution and wavepacket solutions, in Vilenkin wavefunction forms. On the wavepackets, we investigated the expectation value and variance of mini-superspace coordinates and momenta in Section \ref{sect43}. Having constructed bulk semi-classical states, as in \cite{Hartnoll:2022snh,Blacker:2023oan,BlackerNing:2023ezy} we expected the WdW state to be dual to a quantum theory. 
What we do here is specifically identify it as a (1+1)-dimensional $T\bar T$-deformed CFT theory, following \cite{McGough:2016lol}.  
The CFT limit corresponds to the asymptotic boundary of the bulk, and deforming the theory corresponds to moving the theory to effectively living on a radial  slice inside the bulk. 
In Section \ref{sectenergy}, we explicitly matched quantities in the WdW state to quantities in the $T\bar T$ theory. 

Let us also mention some directions for future work. In this work, we have considered a (2+1)-dimensional bulk. The higher-dimensional generalization of $T \bar T$ (in that context called $T^2$) \cite{Hartman:2018tkw,Taylor:2018xcy} should allow a completely equivalent interpretation of the AdS$_4$ solution of \cite{Hartnoll:2022snh} as discussed in our Section \ref{sectenergy}. Inclusion of a gauge field (such as in \cite{BlackerNing:2023ezy}) or inclusion of rotation in such an analysis may also lead to useful insights.  

Connections between $T\bar T$ and WdW have been further explored in the context of Cauchy slice holography \cite{WallAraujo-Regado:2022gvw}. Our work presented here on the \textit{radial} WdW formalism in AdS$_3$ uses a different set-up than the one considered in Cauchy slice holography. Specifically, the $\Sigma$ slices we consider in Figure \ref{figBTZnonrot} are not Cauchy slices anchored on the boundary at constant boundary time. We defer investigations of their set-up to later work. 
In addition, we plan to consider applications of the techniques discussed in this paper to de Sitter spacetimes, where the volume time evolution will correspond to actual timelike evolution. For this, relevant work has very recently appeared \cite{Godet:2024ich}.

\acknowledgments{
	We are grateful to Leonardo Chataignier, Sean Hartnoll, Rodolfo Panerai, Matteo Selle and Ayngaran Thavanesan for helpful discussions. We in particular thank Claus Kiefer for many discussions and shared insights. The research of NC and BH is funded, in part, by the Deutsche Forschungsgemeinschaft (DFG, German Research Foundation) –
	Projektnummer 277101999 – TRR 183 (project A03 and B01). BH acknowledges support by the key profile area Quantum Matter and Materials (QM2) of the University of Cologne.  M.J.B. was supported by a Gates Cambridge Scholarship (OPP1144). SN acknowledges the support from Oxford physics department, New College and London G-Research. NC thanks l'Institut Pascal at Universit\'e Paris-Saclay for hospitality during the completion of this work,  with the support of the program ``Investissements d'avenir'' ANR-11-IDEX-0003-01.  
}

\appendix

\section{ADM deparametrization} \label{AppADM}

The action $I$ of a mechanical system with $M$ true degrees of freedom was discussed in Section \ref{sect2} to have a parameterized incarnation 
\ali{
	I = \int d\tau \left( p_j \p_\tau q_j - N H(p_j,q_j) \right) \qquad (j = 1, ..., M+1). 	
}
This action is invariant under reparametrizations $\tau \ra \tau'(\tau)$ of the parameter $\tau$, with $N(\tau)$ transforming in the same way as $\p_\tau q_j$. Variation with respect to the field $N(\tau)$ imposes the constraint  $H(p_j,q_j) = 0$ with solution $p_{M+1} = -H_{true}(p_i,q_i,q_{M+1})$. It is a Lagrange multiplier, and thus undetermined by the dynamics of the system.  Variation with respect to $p_{M+1}$ sets $q'_{M+1}(\tau) \sim N(\tau)$, 
showing that $q_{M+1}(\tau)$ is equally undetermined by the dynamics as $N(\tau)$. That is, $q_{M+1}(\tau)$ can be freely fixed by a gauge choice or `coordinate condition'. For example, $q_{M+1}(\tau) = \tau$.   

Let us now very explicitly write out the steps that take us from the parameterized form of the action to the unparameterized form. For each step, there will be an explicit equivalent in the gravity discussion. 
\ali{
	I &= \int d\tau \left( p_j \p_\tau q_j - N H(p_j,q_j) \right) \qquad (j = 1, ..., M+1) \label{depstep0} \\ 
	  &= \int d\tau \left. \left( p_j \p_\tau q_j \right)\right.|_{_{p_{M+1} = -H_{true}(p_i,q_i,q_{M+1})}} \qquad (j = 1, ..., M+1, \quad i = 1, ..., M) \label{depstep1} \\ 
	  &= \int d\tau \left( p_i \p_\tau q_i - H_{true}(p_i,q_i,q_{M+1})  \p_\tau q_{M+1} \right) \qquad (i = 1, ..., M) \label{depstep2} \\ 
	  &= \int dq_{M+1} \left( p_i \frac{d q_i}{d q_{M+1}} - H_{true}(p_i,q_i,q_{M+1})  \right) \qquad (i = 1, ..., M) \label{depstep3} \\
	  &= \int dt \left( p_i \frac{d q_i}{d t} - H_{true}(p_i,q_i,t)  \right) \qquad (i = 1, ..., M). \label{depstep4} 
}
The first step, going to \eqref{depstep1}, consists of substituting the constraint solution. In \eqref{depstep2}, the resulting action is written out in terms of the $M$ true variables, and the $(M+1)$'th variables are singled out. 
Next, in \eqref{depstep3}, the gauge choice $q_{M+1}(\tau) = \tau$ is made, and finally in \eqref{depstep4}, the notational choice is made to write $q_{M+1} \equiv t$, identifying a time coordinate $t$ as the independent variable in terms of which the action takes the canonical form \eqref{depstep4}.  
The combination of these last two steps can be referred to as imposing a coordinate condition. It corresponds to the introduction of an `intrinsic' coordinate, as opposed to the arbitrary parameter $\tau$, which is `extrinsic' to the system.

Now we can apply the same steps to the gravitational action 
\ali{
	S &= \int dr \, d^2 x \, \left(\pi^{ij} \p_r \gamma_{ij} - N \mathcal H(\pi^{ij}, \gamma_{ij}) \right) 
}
that we are dealing with in Section \ref{sect2}: 
\ali{
	S &= \int dr \, d^2 x \, \left(\pi^{ij} \p_r \gamma_{ij} - N \mathcal H(\pi^{ij}, \gamma_{ij}) \right)  \\ 
	&= \int dr \, d^2 x \, \left. \left(\pi^{ij} \p_r \gamma_{ij}  \right)\right.|_{_{\pi_v = -\mathcal H_{ADM}(\tilde \pi^{ij},\tilde \gamma_{ij},v)}}  \label{depstep1grav} \\ 
	&= \int dr \, d^2 x \left. \left( \frac{\tilde \pi^{ij}}{v} + \frac{1}{2} \tilde \gamma^{ij} \pi_v \right) \left( v \p_r \tilde \gamma_{ij} + \tilde \gamma_{ij}\p_r v \right) \right|_{_{\pi_v = -\mathcal H_{ADM}(\tilde \pi^{ij},\tilde \gamma_{ij},v)}} \label{depstep2grav} \\
	&= \int dr \, d^2 x \left. \left( \tilde \pi^{ij} \p_r \tilde \gamma_{ij}+ \pi_v \p_r v \right) \right|_{_{\pi_v = -\mathcal H_{ADM}(\tilde \pi^{ij},\tilde \gamma_{ij},v)}} \label{depstep3grav} \\ 
	&= \int dv \, d^2 x  \left( \tilde \pi^{ij} \p_v \tilde \gamma_{ij} - \mathcal H_{ADM}(\tilde \pi^{ij},\tilde \gamma_{ij},v) \right) . \label{depstep4grav}
}
The steps echo the mechanics discussion \eqref{depstep0}-\eqref{depstep4}: the deparametrization consists of first substituting the constraint solution \eqref{ADM416}, with notation \eqref{HADM}, then rewriting in terms of the true degrees of freedom introduced in \eqref{newdof}, and finally imposing 
the coordinate condition $-\sqrt{-\gamma(r)} = r \equiv -v$ that identifies the volume time $v$ as a preferred radial `time'. (For conciseness, we set the AdS radius equal to one in this Appendix.)   

We summarize  
the equivalence of the mechanics and radial gravity discussion in table \ref{table}. 
\begin{table}[t]
	\centering  
	\begin{tabular}{ |c|c|c| } 
		\hline
		 & Mechanics & Radial canonical gravity \\
		\hline
		`time' parameter & $\tau$ & $r$ \\ 
		degrees of freedom & $(p_j,q_j)$ \quad {\scriptsize $(j = 1, ..., M+1)$},  & $(\pi^{ij},\gamma_{ij})$ \\  
		preferred `time'& $q_{M+1}(\tau) = \tau \equiv t$ & $-\sqrt{-\gamma(r)} = r \equiv -v$ \\  
		true Hamiltonian & $p_{M+1}\equiv -H_{true}(p_i,q_i,q_{M+1})$ & $\pi_v \equiv - \mathcal H_{ADM}(\tilde \pi^{ij},\tilde \gamma_{ij},v)$ \\  
		true degrees of freedom & $(p_i,q_i)$ \quad \quad {\scriptsize $(i = 1, ..., M)$} & $(\tilde \pi^{ij},\tilde \gamma_{ij})$ \\  
		\hline
	\end{tabular}
\caption{Comparison of roles in Hamiltonian formulation of classical mechanics and radial canonical AdS$_3$ gravity (for $l=1$).} \label{table}
\end{table}

\section{BTZ from the Hamilton-Jacobi equation in $(v,k)$ notation} \label{Appvol}

This Appendix shows a repetition of the calculations presented in Section \ref{sect3} using a notation that makes the volume dependence explicit. In particular we will use the variables $k$, $v$ and $\omega$. 
The variable $k$ is $\log \tilde \gamma_{tt}$ in relation to the notation in the main part of the paper, $\omega$ is the angular frequency, and $v$ takes on the interpretation of the spatial volume when $\omega$ is zero. 
The metric ansatz for the solution inside the event horizon is given by the following expression: 
\begin{equation}
	ds^2= -N^2 dr^2 + v e^{k} \frac{dt^2}{(\Delta t)^2} + v e^{-k}(\frac{d\varphi}{\Delta \varphi} -\omega \frac{dt}{\Delta t})^2,
	\label{metric-vk}
\end{equation}
where $N, k,v$ and $\omega$ are functions of the radial coordinate $r$, and we have introduced a rescaling of the coordinates so that $v$ is measuring a volume (for $\omega =0$), not a volume density (different from the notation in the main part of the paper). 
Evaluating the gravitational action $S$ on \eqref{metric-vk}, we reach the following Lagrangian:
\begin{equation}
	L= \frac{v}{4 \kappa N}\left((\partial_r k)^2 - \frac{(\partial_r v)^2}{v^2} + e^{-2k}(\partial_r \omega)^2\right) + \frac{N v}{\kappa l^2}.
\end{equation}

From this Lagrangian one can obtain the momenta conjugate to $k,v,\omega$ in the usual way: 
\begin{subequations}
	\begin{align}
		& p_k= \frac{\partial L}{\partial \dot{k}}= \frac{v}{2 \kappa N}\dot{k}; \\
		& p_v= \frac{\partial L}{\partial \dot{v}}=  -\frac{\dot{v}}{2 \kappa N v};  \\
		& p_{\omega}= \frac{\partial L}{\partial \dot{\omega}}=\frac{v}{2\kappa N}e^{-2k}\dot{\omega},
	\end{align}
	\label{momenta-vk}
\end{subequations}
where the dot represents differentiation with respect to $r$. Using these results, the Hamiltonian reads as follows: 
\begin{equation}
	H=  \frac{N \kappa}{v}\left[p_k^2 -v^2 p_v^2+e^{2k}p_{\omega}^2 -\frac{v^2}{\kappa^2 l^2}\right],
	\label{ham-vk}
\end{equation}
and the Hamiltonian constraint: 
\begin{equation}
	-p_k^2 +v^2 p_v^2-e^{2k}p_{\omega}^2 +\frac{v^2}{\kappa^2 l^2}=0,
	\label{ham-cons-vk}
\end{equation}
which yields the following constraint equation for the on-shell action: 
\begin{equation}
	-(\partial_k \mathcal{I})^2+ v^2 (\partial_v \mathcal{I})^2 -j^2 e^{2k} + \frac{v^2}{\kappa^2 l^2}=0,
	\label{cons-vk}
\end{equation}
where we have imposed the condition $\partial_{\omega}\mathcal{I}=j$. This equation is solved by the action given by: 
\begin{equation}
	\mathcal{I}= \frac{v }{\kappa  l}\sinh (k+k_0)-l\, \kappa \frac{j^2 e^{(k-k_0)}}{2 v}+j \omega.
	\label{onshell-vk}
\end{equation}

At this point, one can recover the usual BTZ metric from \eqref{metric-vk}. Firstly, we introduce the constants $\{\omega_0, \epsilon_0\}$ such that: 
\begin{subequations}
	\begin{align}
		\partial_{k_0} \mathcal{I}& = \epsilon_0; \\
		\partial_j \mathcal{I} &= \omega_0.
	\end{align}
	\label{constants}
\end{subequations}
Solving these equations will give expressions for the metric components in terms of $\omega_0$ and $\epsilon_0$. Then, we make the change of variable $v e^{-k}= \frac{l^4}{z^2}$. From the Euler-Lagrange equation for $N$ we derive the following expression: 
\begin{equation}
	N^2 dr^2 = \frac{l^2(v^2(dk^2 + e^{-2k}d\omega^2)-dv^2)}{4 v^2}.
	\label{el-n}
\end{equation}
Making the already mentioned change of variable and inserting the solutions of equations \eqref{constants} into \eqref{el-n}, one reaches the following metric in the usual BTZ form: 
\begin{equation}
	ds^2=\frac{l^2}{z^2}\left(\frac{dz^2}{f(z)} - f(z)d\tilde{t}\,^2 + \left(l d\varphi -\left(e^{k_0}\omega_0 + \frac{j \, \kappa\,  z^2}{l^3}\right) d\tilde{t}\right)^2\right),
\end{equation}
where $z= \frac{l^2}{r}$, $\tilde{t}=  t\, l\, e^{-k_0}$ and
\begin{equation*}
	f(z)= 1- \frac{2 \, \epsilon_0 \, \kappa \, e^{k_0} }{l^3} \, z^2 +\frac{ j^2 \, \kappa^2 }{l^6}z^4 . 
\end{equation*}

This derivation, which is analogous to the one presented in \cite{Hartnoll:2022snh}, shows another way to recover the BTZ solution starting from a mini-superspace ansatz that has an explicit dependence on the volume of the constant $r$ slices ($v$) making it easier to relate it to the volume time discussion in Section \ref{sect2}.

\subsection{Clocks and expectation values}

Following the strategy described in section \ref{sectsc}, the quantized version of the Hamilton-Jacobi equation \eqref{cons-vk} reads as follows: 
\ali{\partial^2_k \Psi- v \p_v(v \p_v \Psi) + \frac{v^2}{\kappa^2 l^2} \Psi =0.\label{wdw-no-rot}}
For simplicity, we have set the angular frequency $\omega=0$. This equation can be solved semiclassically by the one-dimensional WKB approximation 
\ali{\Psi_{\pm}(v, k)= \int \frac{d \epsilon}{2 \pi} \alpha(\epsilon) \psi_{\pm}(v, k; \epsilon),\label{wkb-form}}
with
\ali{\psi_{\pm}(v, k; \epsilon) = \frac{e^{i \epsilon k}}{\left(\epsilon^2 - \frac{v^2}{\kappa^2 l^2 }\right)^{1/4}}\exp\left\{ \pm i \left[ \sqrt{\epsilon^2 - \frac{v^2}{\kappa^2 l^2 }}- \epsilon \tanh^{-1} \frac{\sqrt{\epsilon^2-\frac{v^2}{\kappa^2 l^2 }}}{\epsilon} - \frac{\pi}{4} \right] \right\}.
	\label{wkb-sol}}

In the same way as in the metric components notation, the semiclassical solution is constructed by wavepackets that consist of a Gaussian superposition, in this case of the WKB modes. This is achieved by defining the function $\alpha(\epsilon)$ as: 
\ali{\alpha(\epsilon)= \sqrt{\frac{2 \sqrt{\pi}}{\Delta}} e^{ik_0\epsilon}e^{-(\epsilon- \epsilon_0)^2/(2 \Delta^2)},}
where we need $\Delta \ll 1$ for the wavepacket to be strongly peaked on the classical solution $\epsilon_0$.

Now we can use this result to compute the conserved norm and expectation values. The choice of clock in this case will be the volume $v$ to make contact with Section \ref{sect2}. Although the volume is not a monotonic function in the black hole interior it is still interesting to use it as a clock for the region outside the horizon, in light of the $T\bar T$ interpretation in Section \ref{sect2TTbar}. 
The conserved norm for the $v$ clock is given by 
\ali{|\Psi_\pm|^2_v= \frac{-i}{2}\int dk (\Psi_\pm^* v \p_v \Psi_\pm- \Psi_\pm v \p_v \Psi_\pm^*)= \pm \int \frac{d \epsilon}{2 \pi} |\alpha(\epsilon)|^2.\label{norm-v-clock}}
We make use of this result to compute the expectation value of the momentum $p_v= -i \p_v$, which is the Hamiltonian for the $v$ time. Using the Laplace ordering prescription, and $\Psi_+$ with unit positive norm, the expectation value is given by
\ali{\langle p_v \rangle_v = -\frac12\int dk (\Psi_+^* \p_v(v \p_v \Psi_+)- \p_v\Psi_+ v \p_v \Psi_+^*) = \int \frac{d \epsilon}{2 \pi} |\alpha(\epsilon)|^2 \sqrt{\frac{\epsilon^2}{v^2} - \frac{1}{\kappa^2 l^2 }}.\label{exp-val-piv}} 
In the semiclassical approximation, this is calculated to be 
\ali{\langle p_v\rangle_v= \sqrt{\frac{\epsilon_0^2}{v^2}-\frac{1}{\kappa^2 l^2 }}-\frac{\Delta ^2}{4 \kappa^2 l^2  v^2}\left(\frac{\epsilon_0^2}{v^2}-\frac{1}{\kappa^2 l^2 }\right)^{-3/2}+ \dots\label{exp-val-piv-semi}}
The leading order of this expansion coincides with the classical value of the momentum $p_v= \frac{1}{\kappa \, l} \sinh[k+k0]$. This is calculated from  \eqref{onshell-vk} after substituting the classical value of the volume $v= \epsilon_0 \kappa \, l \, \text{sech}[k+k_0]$.

The quantum variance is given by
\ali{\text{var}(p_v)= \frac{ \Delta ^2 }{2(\epsilon_0^2 \, \kappa^2 l^2- v^2)}+\frac{\Delta ^2}{2 v^2}+\mathcal{O}(\Delta^4). \label{variance-piv}}

\bibliographystyle{JHEP}
\bibliography{referencesWdWDraft}

\end{document}